\documentclass[12pt,draftclsnofoot,onecolumn]{IEEEtran}
\usepackage{pifont}
\usepackage{graphicx}
\usepackage{fancyhdr}
\usepackage{amsmath,amssymb,amstext,amsfonts}
\usepackage{slashbox}
\usepackage{bm}
\usepackage{algorithm}
\usepackage{array}
\usepackage{cite}

\usepackage{subfigure}
\usepackage{balance}
\usepackage{algorithmicx}
\usepackage{algpseudocode}

\usepackage{braket}
\usepackage{tabulary}
\usepackage{diagbox}
\usepackage{amsthm}
\usepackage{color}

\makeatletter
\renewcommand{\maketag@@@}[1]{\hbox{\m@th\normalsize\normalfont#1}}%
\makeatother

\begin{document}

\title{Turbulent Multiple-Scattering Channel Modeling for Ultraviolet Communications: A Monte-Carlo Integration Approach}
\author{Renzhi~Yuan, \IEEEmembership{Member,~IEEE}, Xinyi~Chu, \IEEEmembership{Student Member,~IEEE}, Tao~Shan, Chuang~Yang, and Mugen~Peng, \IEEEmembership{Fellow,~IEEE}.
\thanks{
Renzhi Yuan, Xinyi Chu, Chuang Yang, and Mugen Peng are with the State Key Laboratory of Networking and Switching Technology, Beijing University of Posts and Telecommunications, Beijing, China (e-mails: renzhi.yuan@bupt.edu.cn; xinyi.chu@bupt.edu.cn; chuangyang@bupt.edu.cn; pmg@bupt.edu.cn); Tao Shan is with the School of Engineering, the University of British Columbia, Kelowna, Canada (e-mail: tao.shan@ubc.ca).}
\thanks{Corresponding author: Mugen Peng}
\thanks{This work is supported by the National Natural Science Foundation of China under No. 62201075.}
\thanks{A journal version of this manuscript is under peer-review.}}

\maketitle

\begin{abstract}
Modeling of multiple-scattering channels in atmospheric turbulence is essential for the performance analysis of long-distance non-line-of-sight (NLOS) ultraviolet (UV) communications. Existing works on the turbulent channel modeling for NLOS UV communications either focused on single-scattering cases or estimate the turbulent fluctuation effect in an unreliable way based on Monte-Carlo simulation (MCS) approach. In this paper, we establish a comprehensive turbulent multiple-scattering channel model by using a more efficient Monte-Carlo integration (MCI) approach for NLOS UV communications, where both the scattering, absorption, and turbulence effects are considered. Compared with the MCS approach, the MCI approach is more interpretable for estimating the turbulent fluctuation. To achieve this, we first introduce the scattering, absorption, and turbulence effects for NLOS UV communications in turbulent channels. Then we propose the estimation methods based on MCI approach for estimating both the turbulent fluctuation and the distribution of turbulent fading coefficient. Numerical results demonstrate that the turbulence-induced scattering effect can always be ignored for typical UV communication scenarios. Besides, the turbulent fluctuation will increase as either the communication distance increases or the zenith angle decreases, which is compatible with existing experimental results and also with our experimental results. Moreover, we demonstrate numerically that the distribution of the turbulent fading coefficient for UV multiple-scattering channels under all turbulent conditions can be approximated as log-normal distribution; and we also demonstrate both numerically and experimentally that the turbulent fading can be approximated as a Gaussian distribution under weak turbulence.
\end{abstract}


%
\IEEEpeerreviewmaketitle

\section{Introduction}
\subsection{Background and Motivation}
The ultraviolet (UV) communication \cite{yuan2016review} employing the ``solar-blind" UV signals ($200$-$280$ nm) as the information carriers has attracted increasing attention in recent decades. Due to its inherent advantages including low background noise, high local security, good adaptivity to extreme weather, and ability of non-line-of-sight (NLOS) links, the UV communication becomes a promising technique for communication scenarios where radio-frequency (RF) communications are impermissible or line-of-sight (LOS) links for traditional free-space optical (FSO) are unavailable \cite{yuan2016review}. In recent decade, considerable works have been performed on the NLOS UV channel modeling \cite{yuan2016integral,wu2019single,shen2019modeling,cao2021single,cao2022single,ding2009modeling,drost2011uv,ding2010path,yuan2019importance,yuan2019monte,shan2020modeling}, channel estimation \cite{gong2015channel,gong2016optical,wei2018simultaneous,hu2020non,shen2021lmmse}, coding and modulation \cite{cao2023performance,cao2023power}, diversity reception \cite{qin2017noncoplanar,yuan2023simo,wang2023mimo}, duplex and relay techniques \cite{ardakani2017performance,gong2018full,wang2023nlos}, and experimental tests \cite{sun201771,wang20181mbps,alkhazragi2020gbit}. However, most existing works on the channel modeling ignored the impacts of atmospheric turbulence, which cannot be applied to the turbulent cases under either long communication distances or large elevation angles.

In existing literatures, the major effects of the atmospheric turbulence on the photon propagating process can be divided into two types: the \textit{turbulence-induced scattering effect} when a photon interacts with the atmospheric medium \cite{ishimaru2005theory} and the \textit{turbulent fluctuation effect} when a photon travels in the atmosphere \cite{andrews2005laser}. In existing works, the turbulence-induced scattering effects and the turbulent fluctuation effects were separately studied in the channel modeling of NLOS UV communications. Besides, existing works on turbulent effects in NLOS UV communications either focused on single-scattering cases or estimate the turbulent fluctuation based on a Monte-Carlo simulation (MCS) approach. As it is demonstrated in \cite{yuan2019importance,yuan2019monte}, the Monte-Carlo integration (MCI) approach can provide a more efficient modeling for multiple scattering channels. Besides, as we will demonstrate later in this paper, the MCI approach can also provide a more interpretable logic for estimating the turbulent fluctuation. Therefore, a comprehensive turbulent multiple scattering channel modeling based on MCI approach considering both the scattering, absorption, and turbulence effects  is urgently demanded for long-distance NLOS UV communications.

\subsection{Related Works}

The turbulence-induced scattering effect on light wave propagation in the atmosphere was elaborately studied and summarized in \cite{ishimaru2005theory}, where the turbulent atmosphere is modeled as a random continuum with refractive-index varying randomly and continuously in time and space. Some preliminary results on the receiving power and channel impulse response (CIR) of NLOS optical links were derived in \cite{ishimaru2005theory} based on a single-scattering assumption. The impact of the turbulence-induced scattering effect on the NLOS UV communication was first studied in \cite{xiao2013non}. However, the impact of the turbulence-induced scattering on the phase scattering function for NLOS UV communications was not clearly studied. Besides, a recent study \cite{xu2022improvement} in underwater optical communication also incorporated the turbulence-induced scattering effect into the scattering-based channel model and demonstrated an addition of more than $5$ dB path loss when the turbulence is considered, which seems opposite to the numerical results obtained in \cite{xiao2013non}. Therefore, the impact of the turbulence-induced scattering effect on the NLOS UV communication is still not clearly indicated.

The turbulent fluctuation effect on light wave propagating for LOS optical links was elaborately studied and summarized in \cite{andrews2005laser}. The impact of turbulent fluctuation effect on the NLOS UV communications was first studied in a single-scattering model under small common volume assumption \cite{ding2011turbulence}, where the NLOS link was divided into two LOS links: one from the transmitter to the common volume and another from the common volume to the receiver. When two turbulent fluctuations with log-normal (LN) distribution are introduced into these two LOS links, the authors in \cite{ding2011turbulence} demonstrated that the overall turbulent fluctuation can also be approximated as an LN distribution. Based on \cite{ding2011turbulence}, the bit-error rate performance of NLOS UV communications using this turbulent fluctuation model was studied in \cite{liu2015performance}. Besides, the turbulent single-scattering model in \cite{ding2011turbulence} was extended to narrow beam cases in \cite{shan2020single}. A turbulence-induced attenuation effect for LOS links was introduced into the turbulent channel model of NLOS UV communications in \cite{shan2020single,zuo2012effect,xiao2012non}. However, the turbulence-induced attenuation considered in LOS laser communications comes from the beam wandering and beam spreading effects \cite{kaushal2017free} caused by the misalignment between the transceivers. Therefore, turbulence-induced attenuation cannot directly be applied to NLOS UV links since no alignment is required in NLOS UV communications. Recently, a turbulent channel modeling based on equivalent scattering points approach was proposed in \cite{chu2024turbulent}, where a Gaussian distribution on the turbulent fading coefficient was obtained. However, these works \cite{ding2011turbulence,liu2015performance,shan2020single,zuo2012effect,xiao2012non,chu2024turbulent} are based on the single-scattering model, which cannot be applied to multiple-scattering based long-distance NLOS UV communications.

The turbulent fluctuation model for multiple-scattering was first studied in \cite{wang2013characteristics} based on an MCS approach \cite{drost2011uv}, where a Gamma-Gamma (GG) turbulent fading is introduced into each LOS link between two scatters. However, the authors in \cite{wang2013characteristics} estimated the turbulent fluctuation by simply calculating the variance of the average receiving power. As we will demonstrate in Section \ref{Estimate_Fluctuation}, the attempt to calculate the turbulent fluctuation by estimating the average receiving power of Monte-Carlo based channel models is unreliable due to the properties of Monte-Carlo process. Besides, it was demonstrated that the MCI approaches enjoy more efficient modeling compared with the MCS approaches. Therefore, a more sophisticated turbulent fluctuation model based on the MCI approach is demanded for characterizing the turbulent multiple-scattering channel.

\subsection{Contributions}
Our work aims to establish a turbulent multiple-scattering channel model based on an MCI approach for NLOS UV communications, considering both scattering, absorption, and turbulence effects. To achieve this, we first introduce the scattering, absorption, and turbulence effects of NLOS UV communications in turbulent channels. Then we proposed an improved estimation method based on an MCI approach for both the turbulent fluctuation and the distribution of the turbulent fading coefficient. Both numerical simulation and experimental measurements are conducted to verified our theoretical analysis. Specifically, we summarize the main contributions of this work as follows:
\begin{itemize}
\item We established the first turbulent multiple-scattering channel model based on an MCI approach for NLOS UV communications, where the scattering, absorption, and turbulence effects are considered.
\item We proposed an improved estimation method for turbulent fluctuation and further proposed an estimation method for the distribution of the turbulent fading coefficient for NLOS UV communications.
\item We demonstrated that the turbulence-induced scattering effect can always be ignored for typical UV communication scenarios; and we demonstrated that the turbulent fluctuation for NLOS UV communications will increase as either the communication distance increases or the zenith angle decreases, which is compatible with existing experimental results and also with our experimental results.
\item We demonstrated numerically that the distribution of the turbulent fading coefficient for NLOS UV communications in all turbulent conditions can be approximated by LN distributions; and we also demonstrated both numerically and experimentally that the turbulent fading under weak turbulence can be approximated as a Gaussian distribution.
\end{itemize}

The rest of this paper is organized as follows. We first introduce the photon propagating model of turbulent channels in Section \ref{Atm_Channel}. Based on the photon propagating model, we then derive the receiving power of multiple-scattering turbulent channels in Section \ref{Receiving_Power_Model}. Then we introduce the estimation methods based on an MCI approach for both the average receiving power, the turbulent fluctuation, and the distribution of the turbulent fading coefficient in Section \ref{MCI_model}. We then present some numerical and experimental results in Section \ref{Numerical_Results} and conclude our work in Section \ref{Conclusion}.

\section{Photon Propagating in Turbulent Channels}\label{Atm_Channel}

The non-turbulent atmosphere can be regarded as a group of particles, which can be modeled as randomly distributed scatters including molecules and aerosols \cite{ishimaru2005theory}; whereas the turbulent atmosphere is usually regarded as a group of turbulent eddies according to Kolmogorov's theory of turbulence, which can be modeled as random continuum with refractive-index varying randomly and continuously in time and space \cite{ishimaru2005theory}. Similar to \cite{xu2022improvement}, we consider the scattering effects due to both the random scatterers and the random continuum in this paper. Therefore, the potential scatters include both the particles and the turbulent eddies.

\subsection{Scattering Effects in Turbulent Channels}\label{Scattering_Effects}
\subsubsection{Scattering Effect of Particles}

When the particle size is much smaller than the light wavelength, the scattering process of a light wave can be modeled as a Rayleigh scattering; whereas when the particle size is comparable to or larger than the light wavelength, the scattering process can be modeled as a Mie scattering \cite{yuan2016review}. We denote the Rayleigh scattering coefficient and the Mie scattering coefficient by $k_s^{ray}$ and $k_s^{mie}$, respectively. Then the total scattering coefficient due to the randomly distributed particles is given by
\begin{equation}
k^{par}_s= k_s^{ray}+k_s^{mie}.
\end{equation}

The scattering strength at different directions around the particle is characterized by the phase scattering function \cite{ishimaru2005theory}. For UV signals, the phase scattering functions for Rayleigh scattering and Mie scattering are given by \cite{yuan2016review}
\begin{equation}
\begin{cases}
p^{ray}(\theta_s)\!=\!\frac{3[1+3\gamma+(1-\gamma)\cos^2\theta_s]}{16\pi(1+2\gamma)},\\
p^{mie}(\theta_s)\!=\!\frac{1-g^2}{4\pi}\!\!\left[\frac{1}{(1+g^2\!-\!2g\cos\theta_s)^{{3}/{2}}}\!+\! \frac{f(3\cos^2\theta_s\!-\!1)}{2(1+g^2)^{{3}/{2}}}\right],
\end{cases}
\end{equation}
\noindent where $\theta_s$ is the scattering (zenith) angle between the incident light direction and the scattering light direction; $\gamma$, $g$, and $f$ are the model parameters. The Rayleigh scattering approximates an isotropic scattering and the Mie scattering is a forward direction dominated scattering \cite{yuan2016review}.

\subsubsection{Scattering Effect of Turbulence}
The turbulence-induced scattering effect for a light propagating in a random continuum can be characterized by the differential cross section per unit volume $\sigma(\bm{i},\bm{o})$ \cite{ishimaru2005theory}, where $\bm{i}$ and $\bm{o}$ are the unit vectors for the incident and the scattering directions, respectively. For a statistically homogeneous and isotropic random continuum and a scalar light wave, the differential cross section $\sigma(\bm{i},\bm{o})$ can be obtained as  $\sigma(\bm{i},\bm{o})=2\pi k^4 \Phi_n(k_s)$, where $\Phi_n(k_s)$ is the power spectral density of the turbulence; $k={2\pi}/{\lambda}$ is the wave number of the incident light; $k_s = 2k\sin({\theta_s}/{2})$ and $\theta_s$ is the scattering angle between $\bm{i}$ and $\bm{o}$. We can observe that the differential cross section $\sigma(\bm{i},\bm{o})$ is a function of the relative direction between $\bm{i}$ and $\bm{o}$. Then we can rewrite $\sigma(\bm{i},\bm{o})$ as $\sigma(\theta_s)=2\pi k^4 \Phi_n\left(2k\sin({\theta_s}/{2})\right)$ and further define the phase scattering function due to the turbulence as \cite{ishimaru2005theory}
\begin{equation}\label{phase_scattering_function_def}
\begin{aligned}
p^{tur}(\theta_s)&\triangleq \frac{\sigma(\theta_s)}{\int_{\Omega}\sigma(\theta_s)\mathrm{d}\omega}=\frac{\Phi_n\left(2k\sin({\theta_s}/{2})\right)}{\int_{\theta_s} \Phi_n\left(2k\sin({\theta_s}/{2})\right) \mathrm{d}\theta_s}.
\end{aligned}
\end{equation}

Without loss of generality, we adopt the widely used Booker-Gordon model \cite{ishimaru2005theory,xiao2013non} with $\Phi_n(k_s) =\frac{\langle n_1^2\rangle d_0^3}{\pi^2(1+k_s^2d_0^2)^2}$, where $\langle n_1^2\rangle$ is the variance of refractive index; $d_0$ is the average size of turbulent eddies. When the turbulence is assumed to be isotropic, we have \cite{ishimaru2005theory,xiao2013non} $\langle n_1^2\rangle=\frac{C_n^2L_0^{{2}/{3}}}{1.91}$, where $L_0$ is the outer scale parameter representing the largest turbulent eddy size; $C_n^2$ is the refractive-index structure parameter representing the strength of the turbulence.

Then substituting $\Phi_n(k_s)$ into \eqref{phase_scattering_function_def}, we can obtain the phase scattering function due to turbulence as
\begin{equation}\label{phase_scattering_function_Booker_Gordon}
\begin{aligned}
p^{tur}(\theta_s)=\frac{1+4k^2d_0^2}{4\pi(1+4k^2d_0^2\sin^2({\theta_s}/{2}))^2}.
\end{aligned}
\end{equation}

Because for optical wavebands including UV wavebands, we always have $\lambda\ll d_0$, i.e., $k^2d_0^2 \gg 1$. This indicates that the scattering happens always at forward direction, which is equivalent to a rectilinear propagation effect. Therefore, the turbulence induced scattering effect can be omitted since it has no true effect on the propagating direction of the light.

At last, we can obtain the scattering coefficient in turbulent channels as
\begin{equation}\label{Total_Scattering}
k_s^{tot}=k_s^{par}=k_s^{ray}+k_s^{mie}.
\end{equation}

Similarly, combining the phase scattering functions $p^{ray}(\theta_s)$ and $p^{mie}(\theta_s)$, we can obtain the total phase scattering function in turbulent channel as
\begin{equation}\label{Total_PSH}
p^{tot}(\theta_s)=\frac{k_s^{ray}}{k_s^{tot}}p^{ray}(\theta_s)+\frac{k_s^{mie}}{k_s^{tot}}p^{mie}(\theta_s).
\end{equation}

\subsection{Absorption Effects in Turbulent Channels}\label{Absorption_Effects}
Besides the scattering effects, the light wave can also be selectively absorbed by the particles due to the electronic transition between different energy levels. The absorption coefficient caused by the turbulence depends on the complex dielectric constant $\epsilon=\epsilon_r+i \epsilon_i$, i.e., $k_a^{tur}=k\epsilon_i\epsilon_r^{-\frac{1}{2}}$, where $\epsilon_i=60 \lambda \delta$ and $\delta$ is the atmospheric conductivity. For the atmosphere near the ground, we have $\epsilon_r\approx 1.00059$ and $\delta\approx 2.2\times 10^{-14}$ $\text{S}/\text{m}$. Then we can obtain the absorption coefficient due to the turbulence as $k_a^{tur}\approx 4.1457 \times 10^{-12}$, which is much smaller than the absorption coefficient due to the particles. Therefore, we can always ignore the absorption effect due to the turbulence for UV communications.

Denoting the absorption coefficient due to the particles by $k_a^{par}$, we can obtain the total extinction coefficient of the atmosphere as
\begin{equation}\label{total_extinction}
k^{tot}_e \triangleq k_s^{tot}+k^{par}_a = k_s^{ray}+k_s^{mie}+k^{par}_a.
\end{equation}

\subsection{Fluctuation Effects in Turbulent Channels}\label{Fluctuation_Effects}
Besides the scattering and absorption effects, the turbulence also introduces a fluctuation effect on the light irradiance \cite{andrews2005laser}. The turbulent fluctuation effect can be characterized by the turbulent fading coefficient $\eta$ with $\eta \triangleq P_r/P_t$ and $\eta\geq 0$, where $P_t$ and $P_r$ are the transmit power and receiving power of a traveling process, respectively.

For weak turbulent conditions, the turbulent fading coefficient is usually modeled as an LN distribution with the normalized probability density function (PDF) given by \cite{andrews2005laser}
\begin{equation}\label{eta_pdf_LN}
f_{H,LN}(\eta;d)=\frac{1}{\eta\sqrt{2\pi\sigma_I^2}}\exp\left({-\frac{(\ln \eta+0.5\sigma_I^2(d))^2}{2\sigma_I^2(d)}}\right),
\end{equation}
\noindent where $\sigma_I^2(d)=e^{\sigma_r^2(d)}-1$ and $\sigma_r^2(d)$ is the Rytov variance for a propagating distance $d$. For a plane wave, the Rytov variance is given by \cite{andrews2005laser}
\begin{equation}
\sigma_r^2(d)=1.23C_n^2k^{7/6}d^{11/6}.
\end{equation}

For moderate and strong turbulent conditions, the turbulent fading coefficient is usually modeled as a GG distribution with the normalized PDF given by \cite{andrews2005laser}
\begin{equation}\label{eta_pdf_GG}
\begin{aligned}
f_{H,GG}(\eta;d)&=\frac{2[\alpha(d)\beta(d)]^{\frac{\alpha(d)+\beta(d)}{2}}\eta^{\frac{\alpha(d)+\beta(d)}{2}-1}}{\Gamma(\alpha(d))\Gamma(\beta(d))}\\
&\quad \quad \times K_{\alpha(d)-\beta(d)}\left(2\sqrt{\alpha(d)\beta(d)\eta}\right),\quad \eta\geq 0,
\end{aligned}
\end{equation}
\noindent where $K_v(x)$ is the modified Bessel function of the second kind with parameter $v$; $\alpha(d)$ and $\beta(d)$ are related to the Rytov variance $\sigma_r^2(d)$ as \cite{andrews2005laser}
\begin{equation}
\begin{cases}
\alpha(d)=\left[\exp\left(\frac{0.49\sigma_r^2(d)}{(1+1.11\sigma_r^{12/5})^{7/6}}\right)-1\right]^{-1},\\
\beta(d)=\left[\exp\left(\frac{0.51\sigma_r^2(d)}{(1+0.69\sigma_r^{12/5})^{5/6}}\right)-1\right]^{-1}.
\end{cases}
\end{equation}

For weak turbulence, i.e., $\sigma_r^2 \to 0$, we have $\alpha \approx 1/0.49\sigma_r^2$, $\beta\approx 1/0.51\sigma_r^2$, and the turbulent variance $\text{Var}[\eta] = 1/\alpha+1/\beta+ 1/\alpha\beta \approx \sigma_r^2$. For saturated turbulence, i.e., $\sigma_r^2 \to \infty$, we have $\alpha \to \infty$, $\beta \to 1$, and the turbulent variance $\text{Var}[\eta] \approx 1$.

\section{Receiving Power of Multiple-Scattering in Turbulent Channels}\label{Receiving_Power_Model}

\begin{figure}
\begin{center}
\includegraphics[width=0.48\textwidth, draft=false]{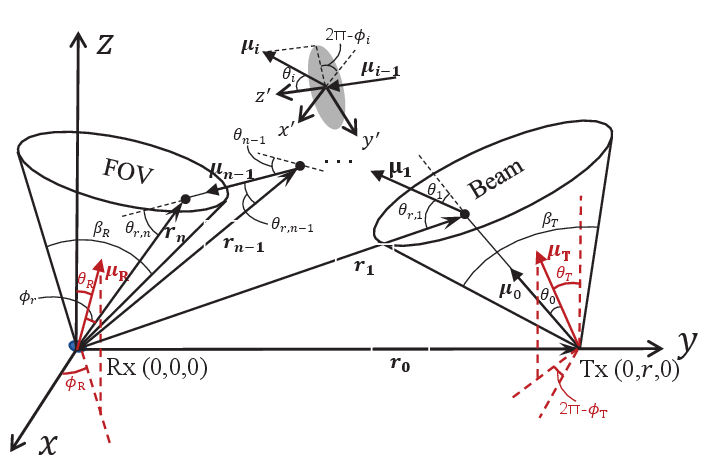}
\caption{Geometry setting for the multiple-scattering process of NLOS UV communications}
\vspace{-0.4cm}
\label{fig:geometry_setting}
\end{center}
\end{figure}

The geometry setting for the multiple-scattering process of NLOS UV communications is shown in Fig. \ref{fig:geometry_setting}. The receiver (Rx) locates at the origin $(0,0,0)$ and the transmitter (Tx) locates at the $y$-axis with coordinates $(0,r,0)$. The direction cosines of Tx and Rx pointing directions are denoted by $\bm{\mu}_T$ and $\bm{\mu}_R$, respectively. The zenith angle and the azimuth angle of the Tx (Rx) are denoted by $\theta_{T(R)}$ and $\phi_{T(R)}$, respectively. The divergence angle of the Tx light beam is denoted by $\beta_T$ and the divergence angle of the Rx field-of-view (FOV) is denoted by $\beta_R$. We regard the initial photon emitting as the $zero$-order scattering and denote the propagating distance, scattering zenith angle and scattering azimuth angle of $i$-order scattering by $d_i$, $\theta_i$ and $\phi_i$, respectively. The position of the $i$th scatter is denoted by $\bm{r}_i$. The direction cosine of $i$-order scattering is denoted by $\bm{\mu}_i$. The initial scattering zenith angle $\theta_0$ is the angle between $\bm{\mu}_0$ and $\bm{\mu}_T$. We consider an $n$-order scattering process and define $\bm{d}\triangleq[d_0, d_1, \cdots, d_{n-1}]^\text{T}$, $\bm{\theta}\triangleq[\theta_0, \theta_1, \cdots, \theta_{n-1}]^\text{T}$, and $\bm{\phi}\triangleq[\phi_0, \phi_1, \cdots, \phi_{n-1}]^\text{T}$.

\subsection{Receiving Power Ignoring Turbulent Fluctuation}
A photon propagating path of an $n$-order scattering is full characterized by the parameters $\{\bm{d},\bm{\theta}, \bm{\phi}\}$. After scattered by the last scatter, the photon can be detected if and only if the position of the $n$th scatter locates in the FOV of the receiver. We denote the set of parameters satisfying this geometrical constrain by $\mathcal{F} \triangleq \{\bm{d},\bm{\theta}, \bm{\phi}|\bm{r}_n \text{ in FOV}\}=\{\bm{d},\bm{\theta}, \bm{\phi}|\bm{r}_n\cdot \bm{\mu}_R \geq  d_n \cos(\beta_R/2)\}$. Then the conditional receiving probability of an $n$-order scattering process for a given propagating path $\{\bm{d},\bm{\theta}, \bm{\phi}\}$ can be approximated as \cite{yuan2019monte}
\begin{equation}\label{conditonal_P_n}
\begin{aligned}
P_n(\bm{d},\bm{\theta}, \bm{\phi})&= I_n(\bm{d},\bm{\theta}, \bm{\phi}) \left(\frac{k_s^{tot}}{k_e^{tot}}\right)^{n} \exp({-k_e^{tot} d_n}) \\
&\quad \quad\times \cos \phi_r \min\left(1, p^{tot}(\theta_{r,n})\Omega_r\right),
\end{aligned}
\end{equation}
\noindent where $d_n\triangleq \|\bm{r}_n\|$ and $\theta_{r,n}$ is the angle between $\bm{\mu}_{n-1}$ and $-\bm{r}_n$; $\phi_r$ is the angle between $\bm{\mu}_R$ and $\bm{r}_n$, as shown in Fig. \ref{fig:geometry_setting}; $\Omega_r$ is the solid angle formed by $\bm{r}_n$ and the receiving area $A_r$; and we have $\Omega_r\approx 2\pi (1-d_n/\sqrt{d_n^2+r_A^2})$, where $r_A=\sqrt{A_r/\pi}$ is the radius of the receiving area; and $I_n(\bm{d},\bm{\theta},\bm{\phi})$ is an indicator function.
The indicator function $I_n(\bm{d},\bm{\theta},\bm{\phi})$ is defined as
\begin{equation}
I_n(\bm{d},\bm{\theta},\bm{\phi}) =
\begin{cases}
1, & \text{$\bm{r_n}$ locates in FOV},\\
0, & \text{otherwise}.
\end{cases}
\end{equation}

Then the conditional receiving probability can be rewritten as
\begin{equation}\label{conditonal_P_n_2}
P_n(\bm{d},\bm{\theta}, \bm{\phi})=
\begin{cases}
P_{d,n}(\bm{d},\bm{\theta}, \bm{\phi}), \quad \{\bm{d},\bm{\theta}, \bm{\phi}\} \in \mathcal{F},\\
0, \quad \{\bm{d},\bm{\theta}, \bm{\phi}\} \notin \mathcal{F},
\end{cases}
\end{equation}
\noindent where $P_{d,n}(\bm{d},\bm{\theta}, \bm{\phi})\triangleq \left(\frac{k_s^{tot}}{k_e^{tot}}\right)^{n} \exp({-k_e^{tot} d_n}) \cos \phi_r \min\left(1, \right.$ $\left.p^{tot}(\theta_{r,n})\Omega_r\right)$.

Then the receiving probability can be obtained by averaging $\{\bm{d},\bm{\theta}, \bm{\phi}\}$ out, i.e.,
\begin{equation}\label{eq:Pr_n_1}
\begin{aligned}
P_n &= \int_{\bm{d}}\int_{\bm{\theta}}\int_{\bm{\phi}}P_n(\bm{d},\bm{\theta}, \bm{\phi})f(\bm{d},\bm{\theta}, \bm{\phi})
\mathrm{d}\bm{d}\mathrm{d}\bm{\theta}\mathrm{d}\bm{\phi},
\end{aligned}
\end{equation}
\noindent where $f(\bm{d},\bm{\theta}, \bm{\phi})=\prod_{i=0}^{n-1}f_D(d_i)f_{\Theta}(\theta_i)f_{\Phi}(\phi_i)$ is the joint PDF of $\{\bm{d},\bm{\theta}, \bm{\phi}\}$. We have used the fact that $\bm{d}$, $\bm{\theta}$, and $\bm{\phi}$ are independent from each other. For the propagating distance $d_i$ with $i\in\{0,1,\cdots,n-1\}$, we have $f_{D}(d_i) = k_e^{tot} e^{-k_e^{tot} d_i}$ with $ 0 \leq d_i\leq \infty$ \cite{yuan2019monte}. For the scattering angle $\theta_i$ with $i=0$, we have $f_{\Theta}(\theta_0)=\frac{\sin\theta_0}{1-\cos({\beta_T}/{2})}$ with $0\leq \theta_0\leq {\beta_T}/{2}$ when a uniform light source is assumed \cite{yuan2019monte}; and for the scattering angle $\theta_i$ with $i\in\{1,2,\cdots,n-1\}$, we have $f_{\Theta}(\theta_i)=2\pi p^{tot}(\theta_i)\sin\theta_i$ with $0\leq \theta_i\leq \pi$ \cite{yuan2019monte}. For the scattering azimuth angle $\phi_i$ with $i\in\{0,1,\cdots,n-1\}$, we have $f_{\Phi}(\phi_i)=\frac{1}{2\pi}$ with $0\leq \phi_i\leq 2\pi$ \cite{yuan2019monte}.

Without loss of generality, we assume that a unit light power is transmitted from the light source. Then the average receiving power equals the receiving probability $P_n$.

\subsection{Receiving Power Considering Turbulent Fluctuation}\label{Receiving_Power_Turbulent}
When the turbulent fluctuation effect is considered, a turbulent fading coefficient is introduced to each photon propagating distance. We denote the turbulent fading coefficient for the $i$th propagating distance $d_i$ by $\eta_i$. Then a photon propagating path $\{\bm{d},\bm{\theta},\bm{\phi}\}$ forms $(n+1)$ propagating distances $\{d_0,d_1,\cdots,d_n\}$, which corresponds to $(n+1)$ turbulent fading coefficients $\{\eta_0,\eta_1,\cdots,\eta_n\}$.

The conditional receiving probability without turbulent fluctuation for a given propagating path $\{\bm{d},\bm{\theta}, \bm{\phi}\}$ is $P_n(\bm{d},\bm{\theta}, \bm{\phi})$ given in \eqref{conditonal_P_n}. Then the conditional receiving probability under turbulent fluctuation effect for a given set of turbulent fading coefficients $\bm{\eta}\triangleq [\eta_0, \eta_1, \cdots,\eta_n]^{\text{T}}$ under a propagating path $\{\bm{d},\bm{\theta}, \bm{\phi}\}$ can be expressed as
\begin{equation}\label{conditional_P_tur_n}
P_{tur,n}(\bm{\eta},\bm{d},\bm{\theta}, \bm{\phi})= P_{n}(\bm{d},\bm{\theta}, \bm{\phi})\prod_{i=0}^n\eta_i .
\end{equation}

Then the receiving probability under turbulent fluctuation for a given propagating path $\{\bm{d},\bm{\theta}, \bm{\phi}\}$ can be obtained by averaging $\bm{\eta}$ out, i.e.,
\begin{equation}\label{eq:Pr_tur_n_1}
\begin{aligned}
P_{tur,n} (\bm{d},\bm{\theta}, \bm{\phi})&=\int_{\bm{\eta}}P_{tur,n}(\bm{\eta},\bm{d},\bm{\theta}, \bm{\phi})f(\bm{\eta}|\bm{d},\bm{\theta},\bm{\phi})\mathrm{d}\bm{\eta},
\end{aligned}
\end{equation}
\noindent where $f(\bm{\eta}|\bm{d},\bm{\theta},\bm{\phi})$ is the joint PDF of $\{\eta_0,\eta_1,\cdots,\eta_n\}$ conditioned on a given propagating path $\{\bm{d},\bm{\theta}, \bm{\phi}\}$. For a given propagating path, $\eta_0, \eta_1, \cdots,\eta_n$ are assumed as independent variables. Then we have
\begin{equation}\label{conditioned_pdf_eta}
\begin{aligned}
f(\bm{\eta}|\bm{d},\bm{\theta},\bm{\phi}) = \prod_{i=0}^{n} f(\eta_i;\sigma_I^2(d_i)),
\end{aligned}
\end{equation}
\noindent where $f(\eta;d)=f_{H,LN}(\eta;d)$ is given in \eqref{eta_pdf_LN} for LN distribution and $f(\eta;d)=f_{H,GG}(\eta;d)$ is given in \eqref{eta_pdf_GG} for GG distribution.

By substituting \eqref{conditional_P_tur_n} and \eqref{conditioned_pdf_eta} into \eqref{eq:Pr_tur_n_1}, we can obtain
\begin{equation}\label{eq:Pr_tur_n_2}
\begin{aligned}
P_{tur,n} (\bm{d},\bm{\theta}, \bm{\phi})
&=P_n(\bm{d},\bm{\theta}, \bm{\phi})\prod_{i=0}^n\left[ \int_{\eta_i}\eta_if(\eta_i;\sigma_I^2(d_i))\mathrm{d}\eta_i\right]\\
&=P_n(\bm{d},\bm{\theta}, \bm{\phi}),
\end{aligned}
\end{equation}
\noindent where we have used the property $ \int_{\eta_i}\eta_if(\eta_i;\sigma_I^2(d_i))\mathrm{d}\eta_i=1$ for $i=0,1,\cdots,n$.

Then the receiving probability considering the turbulent fluctuation effect can be obtained by averaging $\{\bm{d},\bm{\theta}, \bm{\phi}\}$ out, i.e.,
\begin{equation}\label{Pr_tur_n}
\begin{aligned}
P_{tur,n}&=\int_{\bm{d}}\int_{\bm{\theta}}\int_{\bm{\phi}}P_n(\bm{d},\bm{\theta}, \bm{\phi})f(\bm{d},\bm{\theta}, \bm{\phi})
\mathrm{d}\bm{d}\mathrm{d}\bm{\theta}\mathrm{d}\bm{\phi}=P_n.
\end{aligned}
\end{equation}

From \eqref{Pr_tur_n}, we can see that $P_{tur,n} = P_{n}$, which indicates that the average receiving power considering the turbulent fluctuation equals the average receiving power ignoring the turbulent fluctuation. Therefore, in practical implementations, we can estimate the average receiving power by solving the integration in \eqref{eq:Pr_n_1}.

\section{Estimating Turbulent Channels by Using an MCI Approach}\label{MCI_model}
\subsection{Estimating the Average Receiving Power}
The integration in \eqref{eq:Pr_n_1} can be estimated by using an MCI approach. Compared with another widely used MCS approach \cite{ding2009modeling,drost2011uv}, the MCI approach \cite{yuan2019importance,ding2010path,yuan2019monte,shan2020modeling,shen2020improved} enjoys a more flexible sampling function, resulting in a higher computational efficiency. Besides, as we will demonstrate later in Section \ref{Estimate_Fluctuation}, the MCI approach can provide a more interpretable logic for estimating the turbulent fluctuation.

Following the MCI procedure given in \cite{yuan2019monte}, we can rewrite \eqref{eq:Pr_n_1} as
\begin{equation}\label{eq:P_n_for_MCI}
P_{n}=\int_{\bm{d}} \int_{\bm{\theta}} \int_{\bm{\phi}} g_{n}(\bm{d},\bm{\theta},\bm{\phi}) \mathrm{d} \bm{d} \mathrm{d} \bm{\theta} \mathrm{d} \bm{\phi},
\end{equation}
\noindent where $g_{n}(\bm{d},\bm{\theta},\bm{\phi})$ is defined as $g_{n}(\bm{d},\bm{\theta},\bm{\phi}) \triangleq P_n(\bm{d},\bm{\theta},\bm{\phi})f(\bm{d},\bm{\theta}, \bm{\phi})$, and where $P_n(\bm{d},\bm{\theta},\bm{\phi})$ is given in \eqref{conditonal_P_n_2}. Now we can choose a sampling PDF $f_{n}(\bm{d},\bm{\theta},\bm{\phi})$ and further rewrite the receiving power $P_{n}$ in \eqref{eq:P_n_for_MCI} as
\begin{equation}\label{eq:P_n_for_MCI_2}
\begin{aligned}
P_{n}&=\int_{\bm{d}} \int_{\bm{\theta}} \int_{\bm{\phi}} O_{n}(\bm{d},\bm{\theta},\bm{\phi}) f_{n}(\bm{d},\bm{\theta},\bm{\phi}) \mathrm{d} \bm{d} \mathrm{d} \bm{\theta} \mathrm{d} \bm{\phi}  \\
&=\text{E} [O_{n}(\bm{d},\bm{\theta},\bm{\phi})],
\end{aligned}
\end{equation}
\noindent where $O_{n}(\bm{d},\bm{\theta},\bm{\phi})\triangleq g_{n}(\bm{d},\bm{\theta},\bm{\phi})/f_{n}(\bm{d},\bm{\theta},\bm{\phi})$ is called the objective function of MCI approach; and $\text{E}[x]$ denotes the expectation of $x$. Therefore, we have expressed the receiving power $P_{n}$ as the expectation of the objective function $O_{n}(\bm{d},\bm{\theta},\bm{\phi})$ when a sampling PDF $f_{n}(\bm{d},\bm{\theta},\bm{\phi})$ is used to sample $\{\bm{d},\bm{\theta},\bm{\phi}\}$.

Different sampling PDF $f_{n}(\bm{d},\bm{\theta},\bm{\phi})$ can result in different computational efficiency of MCI approach \cite{yuan2019importance,ding2010path,yuan2019monte}. Here we adopt the so-called important sampling method \cite{yuan2019importance}, which enjoys the same convergent speed as the MCS approach but a faster calculating speed. The sampling PDF of the important sampling is given by
\begin{equation}\label{Sampling_function}
f_{n}(\bm{d},\bm{\theta},\bm{\phi})=f(\bm{d},\bm{\theta}, \bm{\phi})= \prod_{i=0}^{n-1} f_D(d_i) f_{\Theta}(\theta_i) f_{\Phi}(\phi_i).
\end{equation}

According to the sampling PDF $f_n(\bm{d},\bm{\theta},\bm{\phi})$ in \eqref{Sampling_function}, we can obtain the sampling functions for generating the propagating distance, the scattering angle, and the scattering azimuth angle as
\begin{equation}\label{sampling_d_i}
d_i=-\frac{\ln (1-\text{rand}(1))}{k_e^{tot}}, \quad i=0, 1, \cdots, n-1,
\end{equation}
\begin{equation}
\theta_i=
\begin{cases}
\arccos\left(1-\text{rand}(1)\times\left(1-\cos(\beta_T/2)\right)\right),\quad i=0,\\
F_{\Theta}^{-1}(\text{rand}(1)), \quad i=1,2,\cdots,n-1,
\end{cases}
\end{equation}
\begin{equation}
\phi_i = 2\pi \times \text{rand}(1), \quad i=0, 1, \cdots, n-1,
\end{equation}
\noindent where $\text{rand}(1)$ is a random number between 0 and 1; $F_{\Theta}^{-1}(\theta_s)$ denotes the inverse cumulative distribution function of the scattering angle $\theta_s$. For $i=1,2,\cdots,n-1$, the scattering angle $\theta_i$ can be solved by using numerical methods, e.g., Newton's bisection search.

Then the corresponding objective function becomes
\begin{equation}\label{Objective_function}
\begin{aligned}
O_{n}(\bm{d},\bm{\theta},\bm{\phi})=
\begin{cases}
P_{d,n}, \quad \{\bm{d},\bm{\theta}, \bm{\phi}\} \in \mathcal{F},\\
0, \quad \{\bm{d},\bm{\theta}, \bm{\phi}\} \notin \mathcal{F}.
\end{cases}
\end{aligned}
\end{equation}

According to the law of large numbers, we can estimate the expectation $P_{n}$ using the average of sampling values for the objective function. Specifically, we can first randomly generate $M$ sampling points $\{\bm{s}_{n}^1,\bm{s}_{n}^2,\cdots,\bm{s}_{n}^M\}$ according to the sampling PDF given in \eqref{Sampling_function}, where we have
\begin{equation}
\bm{s}_{n}^m\triangleq [d_0^m,\theta_0^m,\phi_0^m, \cdots,d_{n-1}^m,\theta_{n-1}^m,\phi_{n-1}^m]^{\text{T}}.
\end{equation}
Then when $M$ is large, we can approximate the receiving power as
\begin{equation}\label{calc_P_tur_n}
P_{n} = \frac {1}{M} \sum_{m=1}^M O_{n}(\bm{s}_{n}^m).
\end{equation}

The total receiving power over $N$ scattering orders can be obtained as
\begin{equation}\label{calc_P_n}
P_{tot}= \sum_{n=1}^N P_{n} = \frac {1}{M} \sum_{m=1}^M \sum_{n=1}^N O_{n}(\bm{s}_n^m).
\end{equation}

\subsection{Estimating the Turbulent Fluctuation}\label{Estimate_Fluctuation}
It is challenging to estimate the turbulent fluctuation of the receiving power by using Monte-Carlo based methods. This is because the Monte-Carlo process itself will introduce a fluctuation effect on the calculated average receiving power \cite{yuan2019monte}. In existing literature \cite{wang2013characteristics}, the turbulent fluctuation was estimated by simply calculating the variance of the average receiving power $P_{tur,n}$, i.e, $\text{Var}[P_{tur,n}]$, obtained from the Monte-Carlo models. However, because $P_{tur,n}=P_{n}$, using the expression in \eqref{calc_P_tur_n}, we can obtain
\begin{equation}\label{variance_MCS}
\text{Var}[P_{tur,n}]=\frac{1}{M^2}\sum_{m=1}^M\text{Var}[O_{n}(\bm{s}_n^m)]=\frac{1}{M}\text{Var}[O_{n}(\bm{d},\bm{\theta},\bm{\phi})],
\end{equation}
\noindent where we have used the fact that all sampling points $\{\bm{s}_n^1, \bm{s}_n^2, \cdots, \bm{s}_n^M\}$ are independent from each other and $\text{Var}[O_{n}(\bm{s}_n^1)]=\text{Var}[O_{n}(\bm{s}_n^2)] \cdots =\text{Var}[O_{n}(\bm{s}_n^M)]=\text{Var}[O_{n}(\bm{d},\bm{\theta},\bm{\phi})]$ is the variance of the objective function.

From eq. \eqref{variance_MCS}, we can see that the variance $\text{Var}[P_{tur,n}]$ depends on both $\text{Var}[O_{n}(\bm{d},\bm{\theta},\bm{\phi})]$ and $M$. Besides, we have $\text{Var}[P_{tur,n}]\to 0$  when the number of sampling points $M \to \infty$, which means the calculated variance can be arbitrarily small as long as $M$ is large enough. This is because a larger number of sampling points $M$ in the Monte-Carlo process will reduce the fluctuation caused by randomness, leading to more stable and precise results. Therefore, the attempt to calculate the turbulent fluctuation by estimating $\text{Var}[P_{tur,n}]$ becomes unreliable due to the fact that $\text{Var}[P_{tur,n}]$ depends on $M$.

Here we present an improved method based on an MCI approach for estimating the turbulent fluctuation, which is more interpretable compared with the existing MCS approach and is independent of $M$ when $M$ is large.

We first substitute \eqref{conditonal_P_n_2} into \eqref{conditional_P_tur_n} and obtain the instantaneous receiving probability as
\begin{equation}\label{conditional_P_tur_n_2}
\begin{aligned}
&P_{tur,n}(\bm{\eta},\bm{d},\bm{\theta}, \bm{\phi})\\
&\quad=
\begin{cases}
P_{d,n}(\bm{d},\bm{\theta}, \bm{\phi})\prod_{i=0}^n\eta_i, \quad \{\bm{d},\bm{\theta}, \bm{\phi}\} \in \mathcal{F},\\
0, \quad \{\bm{d},\bm{\theta}, \bm{\phi}\} \notin \mathcal{F},
\end{cases}
\end{aligned}
\end{equation}
\noindent which can be regarded as a function of random variables $\bm{\eta}$, $\bm{d}$, $\bm{\theta}$, and $\bm{\phi}$. Therefore, the randomness of the receiving power comes from both the turbulent fading $\bm{\eta}$ and the propagating path $\{\bm{d},\bm{\theta},\bm{\phi}\}$. To characterize the turbulent fluctuation effect only, we can average $\{\bm{d},\bm{\theta},\bm{\phi}\}$ out and obtain the instantaneous receiving power conditioned on turbulent fading coefficients as
\begin{equation}\label{P_tur_n_eta}
\begin{aligned}
P_{tur,n}(\bm{\eta})
&=\!\int_{\bm{d}}\int_{\bm{\theta}}\int_{\bm{\phi}} P_{n}(\bm{d},\bm{\theta}, \bm{\phi}) \! \prod_{i=0}^n\eta_i \! f(\bm{d},\bm{\theta}, \bm{\phi})\mathrm{d}\bm{d}\mathrm{d}\bm{\theta}\mathrm{d}\bm{\phi}\\
&=\!{\prod_{i=0}^n\eta_i} P_n.
\end{aligned}
\end{equation}

Then we can define an \textit{equivalent turbulent fading coefficient} as
\begin{equation}
\eta_{eq,n}\triangleq \frac{P_{tur,n}(\bm{\eta})}{P_n}
\end{equation}
\noindent on the average receiving power $P_n$ such that the instantaneous receiving power $P_{tur,n}(\bm{\eta}) = \eta_{eq,n}P_n$. Obviously, we have
\begin{equation}\label{eta_eq_n_def}
\eta_{eq,n}= \prod_{i=0}^n \eta_i.
\end{equation}

One can easily verify that $\text{E}[\eta_{eq,n}]=1$. Then the turbulent fluctuation for the $n$th scattering order can be fully characterized by the turbulent variance
\begin{equation}\label{turbulent_variance}
\begin{aligned}
\sigma_{tur,n}^2&\triangleq \text{Var}[\eta_{eq,n}]=\int_{\bm{\eta}}(\eta_{eq,n}-1)^2f(\bm{\eta})\mathrm{d}\bm{\eta},
\end{aligned}
\end{equation}
\noindent where $f(\bm{\eta})$ is the joint PDF of $\{\eta_0,\eta_1,\cdots,\eta_n\}$. From \eqref{conditional_P_tur_n_2} we can see that those propagating paths with $\{\bm{d},\bm{\theta}, \bm{\phi}\} \notin \mathcal{F}$ contribute zero to the turbulent fluctuation effect. Therefore, we should restrict the variance on the set $\mathcal{F}$. Then the joint PDF $f(\bm{\eta})$ can be obtained by averaging $\{\bm{d},\bm{\theta},\bm{\phi}\}$ out of the conditional PDF $f(\bm{\eta}|\bm{d},\bm{\theta},\bm{\phi})$ in \eqref{conditioned_pdf_eta} on the set $\mathcal{F}$, i.e.,
\begin{equation}\label{unconditioned_pdf_eta}
\begin{aligned}
f_{H}(\bm{\eta})&\!=\!\int_{\mathcal{F}}f(\bm{d},\bm{\theta},\bm{\phi}|\{\bm{d},\bm{\theta},\bm{\phi}\}\in\mathcal{F})\\
&\quad \quad \times f(\bm{\eta}|\bm{d},\bm{\theta},\bm{\phi})\mathrm{d}\bm{d}\mathrm{d}\bm{\theta}\mathrm{d}\bm{\phi},
\end{aligned}
\end{equation}
\noindent \noindent where $f(\bm{d},\bm{\theta},\bm{\phi}|\{\bm{d},\bm{\theta},\bm{\phi}\}\in\mathcal{F})$ is the joint PDF of $\{\bm{d},\bm{\theta},\bm{\phi}\}$ on the set $\mathcal{F}$; and we have
\begin{equation}\label{conditional_PDF_d_theta_phi_on_F}
\begin{aligned}
f(\bm{d},\bm{\theta},\bm{\phi}|\{\bm{d},\bm{\theta},\bm{\phi}\}\in\mathcal{F})
&=\frac{f(\bm{d},\bm{\theta},\bm{\phi})}{\text{Pr}[\{\bm{d},\bm{\theta},\bm{\phi}\}\in\mathcal{F}]},
\end{aligned}
\end{equation}
\noindent where $\text{Pr}[\{\bm{d},\bm{\theta},\bm{\phi}\}\in\mathcal{F}]$ is the probability of $\{\bm{d},\bm{\theta},\bm{\phi}\} \in \mathcal{F}$ given by
\begin{equation}\label{Prob_of_F}
\text{Pr}[\{\bm{d},\bm{\theta},\bm{\phi}\}\in\mathcal{F}]=\int_{\mathcal{F}}f(\bm{d},\bm{\theta},\bm{\phi}) \mathrm{d}\bm{d}\mathrm{d}\bm{\theta}\mathrm{d}\bm{\phi}.
\end{equation}

By substituting \eqref{conditioned_pdf_eta}, \eqref{conditional_PDF_d_theta_phi_on_F}, and \eqref{Prob_of_F} into \eqref{unconditioned_pdf_eta}, we can obtain
\begin{equation}\label{unconditioned_pdf_eta_1}
\begin{aligned}
f_{H}(\bm{\eta})&=\frac{\int_{\mathcal{F}}\prod_{i=0}^{n} f(\eta_i;d_i)
f(\bm{d},\bm{\theta},\bm{\phi})
\mathrm{d}\bm{d}\mathrm{d}\bm{\theta}\mathrm{d}\bm{\phi}}
{\int_{\mathcal{F}}f(\bm{d},\bm{\theta},\bm{\phi}) \mathrm{d}\bm{d}\mathrm{d}\bm{\theta}\mathrm{d}\bm{\phi}}\\
&=\frac{\int_{\bm{d}}\int_{\bm{\theta}}\int_{\bm{\phi}}\prod_{i=0}^{n} f(\eta_i;d_i)I_n
f(\bm{d},\bm{\theta},\bm{\phi})
\mathrm{d}\bm{d}\mathrm{d}\bm{\theta}\mathrm{d}\bm{\phi}}
{\int_{\bm{d}}\int_{\bm{\theta}}\int_{\bm{\phi}}
I_nf(\bm{d},\bm{\theta},\bm{\phi}) \mathrm{d}\bm{d}\mathrm{d}\bm{\theta}\mathrm{d}\bm{\phi}}.
\end{aligned}
\end{equation}

Then by substituting \eqref{eta_eq_n_def} and \eqref{unconditioned_pdf_eta_1} into \eqref{turbulent_variance}, we can obtain the turbulent variance as
\begin{equation}\label{sigma_tur_n_2}
\begin{aligned}
\sigma_{tur,n}^2 &\!=\!\frac{\int_{\bm{d}}\int_{\bm{\theta}}\int_{\bm{\phi}}
(\prod_{i=0}^n M_2(d_i)-1)
I_nf(\bm{d},\bm{\theta},\bm{\phi})
\mathrm{d}\bm{d}\mathrm{d}\bm{\theta}\mathrm{d}\bm{\phi}}
{\int_{\bm{d}}\int_{\bm{\theta}}\int_{\bm{\phi}}
I_n f(\bm{d},\bm{\theta},\bm{\phi})
\mathrm{d}\bm{d}\mathrm{d}\bm{\theta}\mathrm{d}\bm{\phi}},
\end{aligned}
\end{equation}
\noindent where $M_2(d_i)$ is the second-order moment of the turbulent fading coefficient $\eta_i(d_i)$. For LN distribution, we have $M_2(d_i)=\exp(\sigma_r^2(d_i))$; and for GG distribution, we have $M_2(d_i)=(1+1/\alpha(d_i))(1+1/\beta(d_i))$.

Similar to the estimation of the average receiving power $P_n$, we can also use the MCI approach to estimate both the nominator and the denominator of $\sigma_{tur,n}^2$ in \eqref{sigma_tur_n_2} using the same sampling PDF given in \eqref{Sampling_function}. Then using the same sampling points $\{\bm{s}_n^1, \bm{s}_n^2, \cdots, \bm{s}_n^M\}$, we can estimate the turbulent variance $\sigma_{tur,n}^2$ as
\begin{equation}\label{estimate_sigma_tur_n_2}
\begin{aligned}
\sigma_{tur,n}^2&=\frac{\frac{1}{M}\sum_{m=1}^M \left[\prod_{i=0}^n M_2(d_i^m)-1\right]I_n(\bm{s}_n^m)}{\frac{1}{M}\sum_{m=0}^M I_n(\bm{s}_n^m)}\\
&=\frac{1}{Count}\sum_{m=1}^M \left[\prod_{i=0}^n M_2(d_i^m)-1\right]I_n(\bm{s}_n^m),
\end{aligned}
\end{equation}
\noindent where $Count \triangleq \sum_{m=0}^M I_n(\bm{s}_n^m)$ is the number of sampling points satisfying $\{\bm{d},\bm{\theta},\bm{\phi}\} \in \mathcal{F}$.

Similar to the $n$th scattering order, the total turbulent fluctuation effect on the total receiving power of $N$ scattering orders can be characterized by defining an equivalent turbulent fading coefficient as
\begin{equation}
\eta_{eq} \triangleq \frac{\sum_{n=1}^N P_{tur,n}(\bm{\eta})}{\sum_{n=1}^N P_n},
\end{equation}
which allows us to treat the complex multiple-scattering channel as a single equivalent link with a unified fading parameter and simplify the characterization of turbulent fluctuation effects. Using the relations $ P_{tur,n}(\bm{\eta})=\eta_{eq,n}P_n$ and $P_{tot}=\sum_{n=1}^N P_n$, we can obtain
\begin{equation}\label{eta_eq_def}
\begin{aligned}
\eta_{eq} &=\sum_{n=1}^N \frac{P_n}{P_{tot}}\eta_{eq,n}.
\end{aligned}
\end{equation}
Then the total turbulent variance $\sigma_{tur}^2$ for $N$ scattering orders can be obtained as
\begin{equation}
\begin{aligned}
\sigma_{tur}^2&\triangleq \text{Var}[\eta_{eq}]=\sum_{n=1}^N \left(\frac{P_n}{P_{tot}}\right)^2\sigma_{tur,n}^2,
\end{aligned}
\end{equation}
\noindent where $P_n$, $P_{tot}$, and $\sigma_{tur,n}^2$ are given by \eqref{calc_P_tur_n}, \eqref{calc_P_n}, and \eqref{estimate_sigma_tur_n_2}, respectively.

\subsection{Estimating the Distribution of Turbulent Fading Coefficient}
Now we further derive the distributions of the equivalent turbulent fading coefficients $\eta_{eq,n}$ and $\eta_{eq}$. For a given propagating path $\{\bm{d},\bm{\theta},\bm{\phi}\}$, the equivalent turbulent fading coefficient $\eta_{eq,n}$ can be obtained from \eqref{eta_eq_n_def} as
\begin{equation}\label{condtional_eta_eq_n}
\eta_{eq,n}(\bm{d},\bm{\theta},\bm{\phi})=\prod_{i=0}^n \eta_{i}(\bm{d},\bm{\theta},\bm{\phi}).
\end{equation}

The logarithm of $\eta_{eq,n}(\bm{d},\bm{\theta},\bm{\phi})$ is
\begin{equation}\label{log_eta_eq_n}
\ln \eta_{eq,n}(\bm{d},\bm{\theta},\bm{\phi})=\sum_{i=0}^n \ln \eta_{i}(\bm{d},\bm{\theta},\bm{\phi}).
\end{equation}

When $\eta_{i}(\bm{d},\bm{\theta},\bm{\phi})$ satisfies an LN distribution with parameter $\sigma_I^2(d_i)$ and $\eta_0,\eta_1,\cdots,\eta_n$ are independent variables, from \eqref{log_eta_eq_n} we can see that $\eta_{eq,n}(\bm{d},\bm{\theta},\bm{\phi})$ also satisfies an LN distribution with parameter $\sigma_{eq,n}^2(\bm{d},\bm{\theta},\bm{\phi}) = \sum_{i=0}^n \sigma_I^2(d_i)$. Therefore, the conditional PDF of $\eta_{eq,n}(\bm{d},\bm{\theta},\bm{\phi})$ is given by
\begin{equation}\label{conditional_PDF_eta_eq_n}
\begin{aligned}
f_{H,n}(\eta_{eq,n}|\bm{d},\bm{\theta},\bm{\phi})&=
\frac{\exp\left(\!{-\frac{(\ln \eta_{eq,n}+0.5\sigma_{eq,n}^2(\bm{d},\bm{\theta},\bm{\phi}))^2}{2\sigma_{eq,n}^2(\bm{d},\bm{\theta},\bm{\phi})}}\!\right)}{\eta_{eq,n}\sqrt{2\pi\sigma_{eq,n}^2(\bm{d},\bm{\theta},\bm{\phi})}}.
\end{aligned}
\end{equation}

Then the unconditional PDF for the equivalent turbulent fading coefficient $\eta_{eq,n}$ can be obtained by averaging $\{\bm{d},\bm{\theta},\bm{\phi}\}$ on $\mathcal{F}$ out. Specifically, we have
\begin{equation}\label{unconditional_PDF_eta_eq_n_0}
\begin{aligned}
f_{H,n}(\eta_{eq,n})&=\int_{\mathcal{F}}f(\bm{d},\bm{\theta},\bm{\phi}|\{\bm{d},\bm{\theta},\bm{\phi}\}\in\mathcal{F})\\
&\quad \quad \times f_{H,n}(\eta_{eq,n}|\bm{d},\bm{\theta},\bm{\phi})\mathrm{d}\bm{d}\mathrm{d}\bm{\theta}\mathrm{d}\bm{\phi},
\end{aligned}
\end{equation}
\noindent where $f(\bm{d},\bm{\theta},\bm{\phi}|\{\bm{d},\bm{\theta},\bm{\phi}\}\in\mathcal{F})$ is given in \eqref{conditional_PDF_d_theta_phi_on_F}. By submitting eqs. \eqref{conditional_PDF_d_theta_phi_on_F} and \eqref{Prob_of_F} into \eqref{unconditional_PDF_eta_eq_n_0}, we can obtain
\begin{equation}\label{unconditional_PDF_eta_eq_n_1}
\begin{aligned}
&f_{H,n}(\eta_{eq,n})\\
&\quad=\frac{\int_{\mathcal{F}}f(\bm{d},\bm{\theta},\bm{\phi}) f_{H,n}(\eta_{eq,n}|\bm{d},\bm{\theta},\bm{\phi})\mathrm{d}\bm{d}\mathrm{d}\bm{\theta}\mathrm{d}\bm{\phi}}{\int_{\mathcal{F}}f(\bm{d},\bm{\theta},\bm{\phi}) \mathrm{d}\bm{d}\mathrm{d}\bm{\theta}\mathrm{d}\bm{\phi}}\\
&\quad=\frac{\int_{\bm{d}}\int_{\bm{\theta}}\int_{\bm{\phi}}
f_{H,n}(\eta_{eq,n}|\bm{d},\bm{\theta},\bm{\phi})I_n f(\bm{d},\bm{\theta},\bm{\phi})
\mathrm{d}\bm{d} \mathrm{d}\bm{\theta} \mathrm{d}\bm{\phi}}
{\int_{\bm{d}}\int_{\bm{\theta}}\int_{\bm{\phi}}
I_nf(\bm{d},\bm{\theta},\bm{\phi})
\mathrm{d}\bm{d}\mathrm{d}\bm{\theta}\mathrm{d}\bm{\phi}},
\end{aligned}
\end{equation}
\noindent where we have used the equality $\int_{\mathcal{F}}g(\bm{d},\bm{\theta},\bm{\phi})f(\bm{d},\bm{\theta},\bm{\phi})$ $\mathrm{d}\bm{d}\mathrm{d}\bm{\theta}\mathrm{d}\bm{\phi}=\int_{\bm{d}}\int_{\bm{\theta}}\int_{\bm{\phi}} I_n g(\bm{d},\bm{\theta},\bm{\phi})f(\bm{d},\bm{\theta},\bm{\phi})\mathrm{d}\bm{d}\mathrm{d}\bm{\theta}\mathrm{d}\bm{\phi}$ for any function $g(\bm{d},\bm{\theta},\bm{\phi})$. This is because $I_n=0$ if $(\bm{d},\bm{\theta},\bm{\phi})$ is not in $\mathcal{F}$. Therefore, by substituting \eqref{conditional_PDF_eta_eq_n} into \eqref{unconditional_PDF_eta_eq_n_1}, we can finally obtain the unconditional PDF for the equivalent turbulent fading coefficient $\eta_{eq,n}$, which is given in \eqref{unconditional_PDF_eta_eq_n} on the top of next page.

\begin{figure*}
\begin{equation}\label{unconditional_PDF_eta_eq_n}
\begin{aligned}
f_{H,n}(\eta_{eq,n})&=\frac{\int_{\bm{d}}\int_{\bm{\theta}}\int_{\bm{\phi}}
\frac{I_n}{\eta_{eq,n}\sqrt{2\pi\sum_{i=0}^n \sigma_I^2(d_i)}}\exp\left({-\frac{(\ln \eta_{eq,n}+0.5\sum_{i=0}^n \sigma_I^2(d_i))^2}{2\sum_{i=0}^n \sigma_I^2(d_i)}}\right)
f(\bm{d},\bm{\theta},\bm{\phi})
\mathrm{d}\bm{d} \mathrm{d}\bm{\theta} \mathrm{d}\bm{\phi}}
{\int_{\bm{d}}\int_{\bm{\theta}}\int_{\bm{\phi}}
I_nf(\bm{d},\bm{\theta},\bm{\phi})
\mathrm{d}\bm{d}\mathrm{d}\bm{\theta}\mathrm{d}\bm{\phi}}.
\end{aligned}
\end{equation}
\end{figure*}

Then we can also adopt the MCI approach to estimate $f_{H,n}(\eta_{eq,n})$ by using the same sampling PDF given in \eqref{Sampling_function}. Using the same sampling points $\{\bm{s}_n^1, \bm{s}_n^2, \cdots, \bm{s}_n^M\}$, we can estimate the PDF of equivalent turbulent fading coefficient $f_{n}(\eta_{eq,n})$ as
\begin{equation}\label{estimate_f_eta_eq_n}
\begin{aligned}
f_{H,n}(\eta_{eq,n})&=\frac{1}{Count}\sum_{m=1}^M \frac{I_n(\bm{s}_n^m)}{\eta_{eq,n}\sqrt{2\pi\sum_{i=0}^n \sigma_I^2(d_i^m)}}\\
&\times \exp\left({-\frac{(\ln \eta_{eq,n}+0.5\sum_{i=0}^n \sigma_I^2(d_i^m))^2}{2\sum_{i=0}^n \sigma_I^2(d_i^m)}}\right).
\end{aligned}
\end{equation}

According to \eqref{eta_eq_def}, the PDF of the total equivalent turbulent fading coefficient $\eta_{eq}$ for $N$ scattering orders can be obtained as the convolution of the PDFs of $\{\frac{P_1}{P_{tot}}\eta_{eq,1},\frac{P_2}{P_{tot}}\eta_{eq,2},\cdots,\frac{P_N}{P_{tot}}\eta_{eq,N}\}$, i.e.,
\begin{equation}
\begin{aligned}
f_{H}(\eta_{eq})&= \frac{P_{tot}}{P_1} f_{H,1}\left(\frac{P_{tot}}{P_1}\eta_{eq}\right)*\frac{P_{tot}}{P_2} f_{H,2}\left(\frac{P_{tot}}{P_2}\eta_{eq}\right)\\
&\quad *\cdots*\frac{P_{tot}}{P_N} f_{H,N}\left(\frac{P_{tot}}{P_N}\eta_{eq}\right),
\end{aligned}
\end{equation}
\noindent where $*$ denotes the convolution operation and $f_{H,n}(x)$ is calculated by \eqref{estimate_f_eta_eq_n}.

The pseudocode of the MCI approach is similar to the one we presented in \cite{yuan2019monte}. The only difference is that we need to add the estimation of turbulent fluctuation and distribution in each MCI process according to the equations given in \eqref{estimate_sigma_tur_n_2} and \eqref{estimate_f_eta_eq_n}, respectively.

We remark that though our derivation here is based on LN distribution for weak turbulence, we can also extend this derivation to moderate and strong turbulence cases. However, in moderate and strong turbulence, a large $C_n^2$ can result in a large Rytov variance $\sigma_r^2$ and thus a large turbulent variance $\text{Var}[\eta]\gg 1$ if LN distribution is adopted, which does not match practical moderate and strong turbulence with $\text{Var}[\eta]$ normally less than 2. An alternative way to extend our method to moderate and strong turbulence is using the Gamma-Gamma fading to calculate the turbulent variance $\text{Var}[\eta]$ since it never can never exceed 2; then we use this turbulent variance $\text{Var}[\eta]$ to calculate the parameter $\sigma_I^2$ in LN fading as $\sigma_I^2=\ln(\text{Var}[\eta]+1)$. Through this way we can estimate the distribution of turbulent fading coefficient for all turbulent conditions.

\section{Simulation and Experimental Results}\label{Numerical_Results}
In this section, we present some simulation and experimental results to verify our theoretical analysis. Unless otherwise specified, the simulation parameters we used in this section are listed in Table \ref{Simulation_Parameters}. Without loss of generality, we set $C_n^2 = 10^{-17}$ $\text{m}^{-2/3}$, $C_n^2=10^{-15}$ $\text{m}^{-2/3}$, and $C_n^2=10^{-13}$ $\text{m}^{-2/3}$ for weak, moderate, and strong turbulent conditions, respectively.

\begin{table}[!t]
\caption{Simulation Parameters}
\label{Simulation_Parameters}
\centering
\begin{tabular}{p{0.2\textwidth}p{0.25\textwidth}}
\hline
\hline
\textbf{Parameters} & \textbf{Values} \\
\hline
$r$ & $500$ m \\
$[\theta_T, \phi_T]$ & $[45^{\circ}, -90^{\circ}]$ \\
$[\theta_R, \phi_R]$ & $[45^{\circ}, 90^{\circ}]$ \\
$[\beta_T, \beta_R]$ & $[17^{\circ}, 30^{\circ}]$\\
$A_r$ & $1.77\times10^{-4} { \text{m}}^{2}$ \\
$\lambda$ & $2.6 \times 10^{-7}$ m \\
$c$  & $2.998\times 10^{8}$ $\text{m}/\text{s}$ \\
$[k_a^{par}, k_s^{ray}, k_s^{mie}]$ & $[0.802, 0.266, 0.284]$ ${\text{km}}^{-1}$  \\
$[\gamma, g, f]$ & $[0.017, 0.72, 0.5]$  \\
$L_0$ & $10^2$ m \\
$d_0$ & $10^{-3}$ m \\
$C_n^2$ & $10^{-15}$ $\text{m}^{-2/3}$ \\
$\Delta t$ & $0.02$ $\mu$s \\
$M$ & $10^6$ \\
$N$ & $3$ \\
\hline
\hline
\end{tabular}
\end{table}

\subsection{Simulation Results}

\begin{figure}
\begin{center}
\includegraphics[width=0.45\textwidth, draft=false]{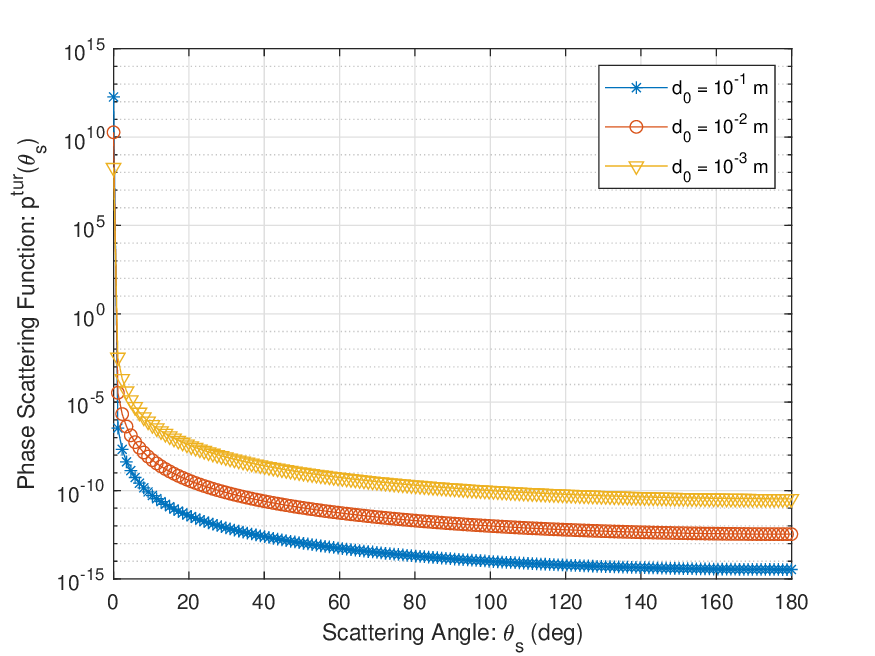}
\caption{Phase scattering function due to turbulence in different $d_0$}
\label{P_tur_theta_s}
\end{center}
\end{figure}

We first explore the turbulence-induced phase scattering function $p^{tur}(\theta_s)$ under different scattering angle $\theta_s$ and correlation distance $d_0$ in Fig. \ref{P_tur_theta_s}. Note that $L_0$ and $C_n^2$ have no impact on the phase scattering function. From Fig. \ref{P_tur_theta_s}, we can see that, the scattering intensity rapidly increases as the scattering angle decreases. This indicates that the incident photons almost always maintain their original traveling direction after scattering since almost all the scattering intensity concentrates at $\theta_s = 0^{\circ}$. In this case, the overall turbulence-induced scattering can be equivalently regarded as a rectilinear propagation with negligible divergence angle, which means no evident impact on the final receiving power can be observed. Therefore, we can ignore the turbulence-induced scattering effect in UV wavebands.

\begin{figure}
\centering
\subfigure[\label{var_tur_vs_M_wangpeng}]
{\includegraphics[width=0.45\textwidth]{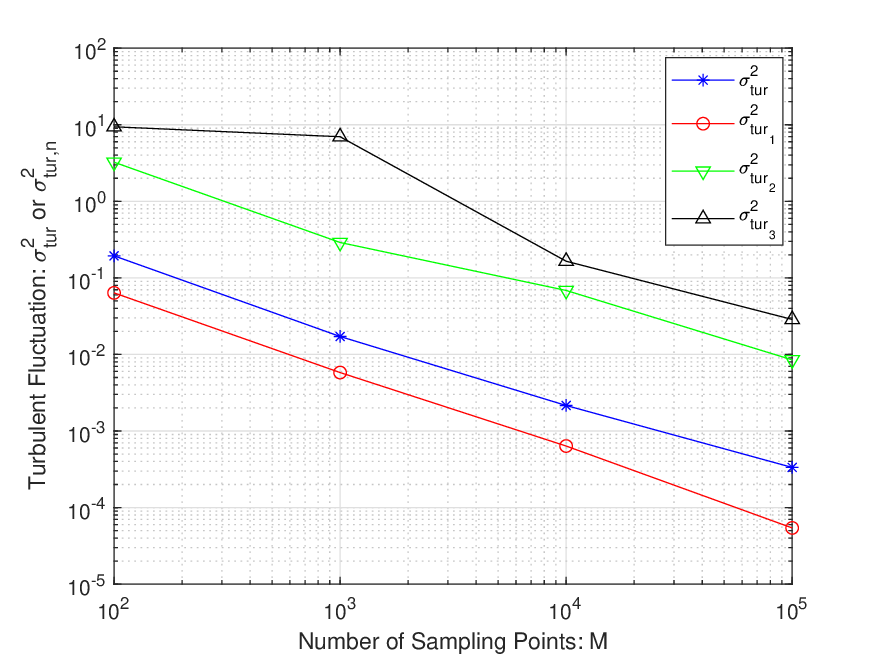}}
\subfigure[\label{var_tur_vs_M_model}]
{\includegraphics[width=0.45\textwidth]{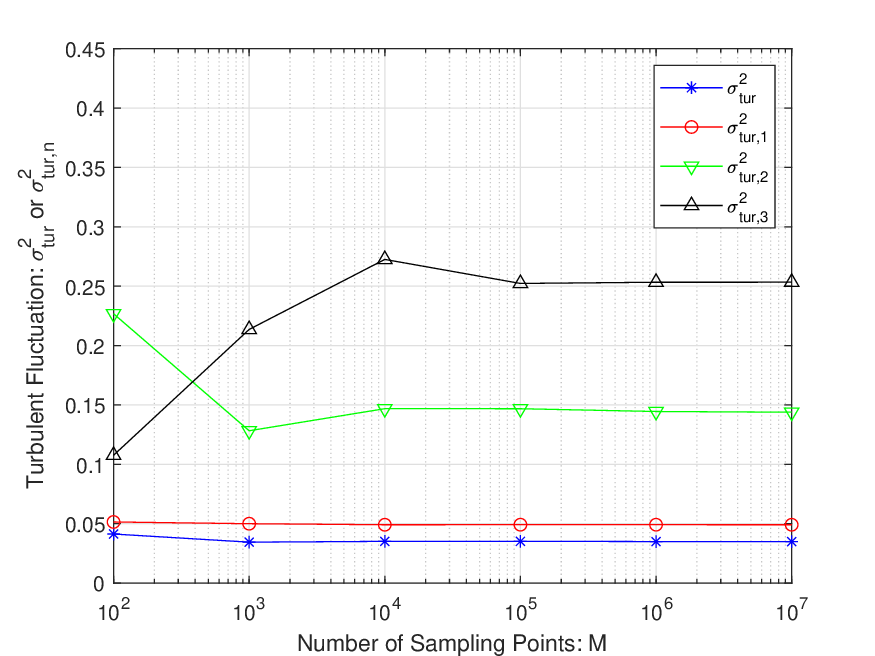}}
\caption{Turbulent fluctuations under different number of sampling points: (a) conventional MC method in \cite{wang2013characteristics}; (b) improved MCI method}
\label{var_tur_vs_M}
\end{figure}

Then we explore the turbulent fluctuation effects. Fig. \ref{var_tur_vs_M} presents the estimated turbulent fluctuation under different number of sampling points $M$. To numerically demonstrate the necessity of our model, we first use the conventional Monte-Carlo method used in \cite{wang2013characteristics}. The simulation results, obtained under varying $M$ with 100 realizations, are presented in Fig. \ref{var_tur_vs_M_wangpeng}. Consistent with the prediction from Eq. \eqref{variance_MCS}, the turbulent fluctuation estimated by the conventional Monte-Carlo method exhibits a decrease as $M$ increases from $10^2$ to $10^5$. The dependency on $M$ indicates that the results are dominated by Monte-Carlo sampling variance rather than physical turbulence. In contrast, our improved MCI method provides stable values for turbulent variance, as shown in Fig. \ref{var_tur_vs_M}(b), where the estimated turbulent fluctuation $\sigma^2_{tur}$ or $\sigma_{tur,n}^2$ becomes stable as $M$ increases. Besides, we can also roughly estimate the turbulent variance by using the method proposed for the single-scattering link in \cite{ding2011turbulence}. Specifically, the turbulent variance can be approximated as $\sigma_{tur,1}^2 \approx \exp(\sigma_{I}^2(r_1)+\sigma_{I}^2(r_2))-1 \approx 0.0493$, which is quite close to our result $\sigma_{tur,1}^2 =0.0492$ in Fig. \ref{var_tur_vs_M}. This verified that our proposed estimation method for turbulent fluctuation can well estimate the turbulent variance.

\begin{figure}
\centering
\subfigure[\label{var_tur_vs_r}]
{\includegraphics[width=0.45\textwidth]{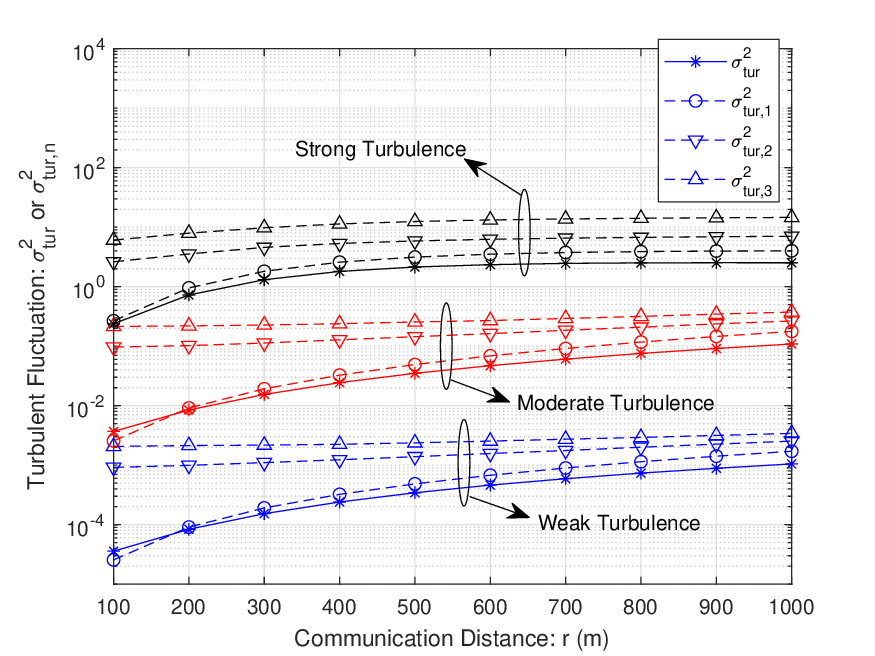}}
\subfigure[\label{var_tur_vs_theta_T}]
{\includegraphics[width=0.45\textwidth]{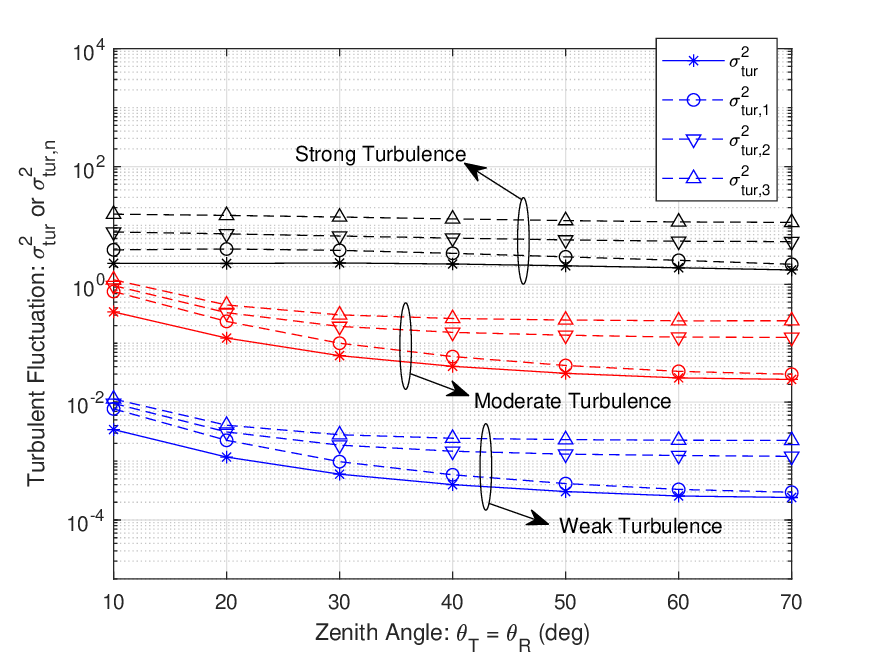}}
\caption{Turbulent fluctuations under various system geometries: (a) turbulent fluctuations under different communication distances; (b) turbulent fluctuations under different zenith angles}\label{fig: var_tur_vs_geometry}
\end{figure}

Then we present the turbulent fluctuations results under various system geometries in Fig. \ref{fig: var_tur_vs_geometry}. From Figs. \ref{var_tur_vs_r} and \ref{var_tur_vs_theta_T}, we can see that the turbulent variance increases as either the communication distance $r$ increases or the zenith angle $\theta_{T}$ decreases \footnote{This indicates that a worse bit-error rate (BER) performance will be obtained when the communication distance or the elevation angle increases, which is compatible with the results in \cite{he2024demonstration}.}, which is compatible with reported experimental results in \cite{chen2014experimental,liao2014turbulence}. Besides, we can also observe that the turbulence variance will approach a stable value for strong turbulent conditions in Figs. \ref{var_tur_vs_r} and \ref{var_tur_vs_theta_T}. This is because the channel will become a saturated turbulent channel under strong turbulence as either the communication distance $r$ increases or the zenith angle $\theta_{T}$ decreases, which is compatible with the saturated turbulence case with $\sigma_r^2 \to \infty$.

\begin{figure}
\centering
\subfigure[\label{var_com_ding}]
{\includegraphics[width=0.45\textwidth]{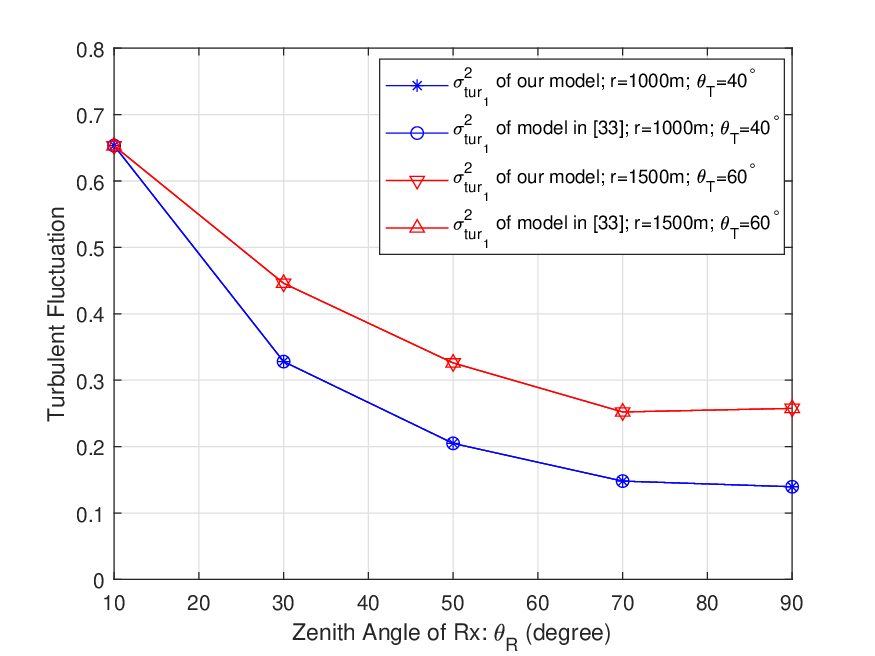}}
\subfigure[\label{var_com_experiment}]
{\includegraphics[width=0.45\textwidth]{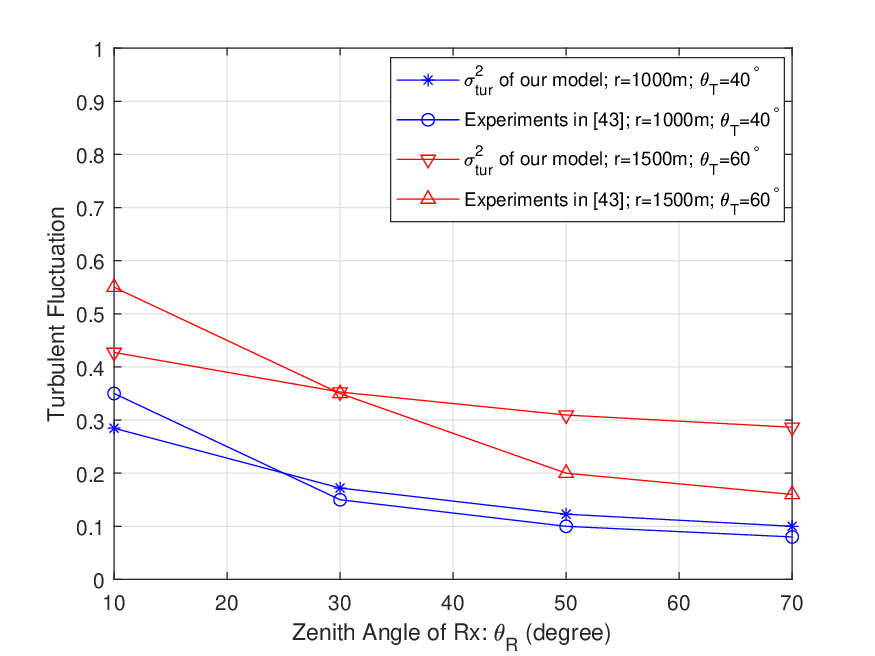}}
\caption{Comparison of the proposed model with existing related works \cite{ding2011turbulence} and \cite{chen2014experimental}: (a) Comparison of single-scattering turbulent fluctuation with \cite{ding2011turbulence}; (b) Comparison of total turbulent fluctuation with experimental results in \cite{chen2014experimental}.}
\label{fig:var_comparison}
\end{figure}

To further validate the accuracy of the proposed model, we compare our results with the single-scattering turbulence model in \cite{ding2011turbulence} and the experimental data reported in \cite{chen2014experimental}, respectively, as shown in Fig. \ref{fig:var_comparison}. The simulation parameters are set according to the experimental configuration in \cite{chen2014experimental}, with the refractive-index structure parameter remains $C_n^2 = 10^{-15} \text{m}^{-2/3}$ and $\theta_R=40^\circ$. Two specific geometric configurations were selected for this comparison: (i) $r=1000$ m with $\theta_T=40^\circ$, and (ii) $r=1500$ m with $\theta_T=60^\circ$. Fig. \ref{var_com_ding} compares the single-scattering turbulent fluctuation $\sigma^2_{tur,1}$ calculated by our model against the analytical model in \cite{ding2011turbulence}. It can be observed that our results highly consist with the existing analytical model across different geometric configurations. Furthermore, Fig. \ref{var_com_experiment} compares the total turbulent fluctuation $\sigma^2_{tur}$ of our multiple-scattering model with the experimental measurements from \cite{chen2014experimental}. The results indicate that our model fits the experimental data well.

\begin{figure*}
\centering
\subfigure[\label{PDF_weak_tur}]
{\includegraphics[width=0.45\textwidth]{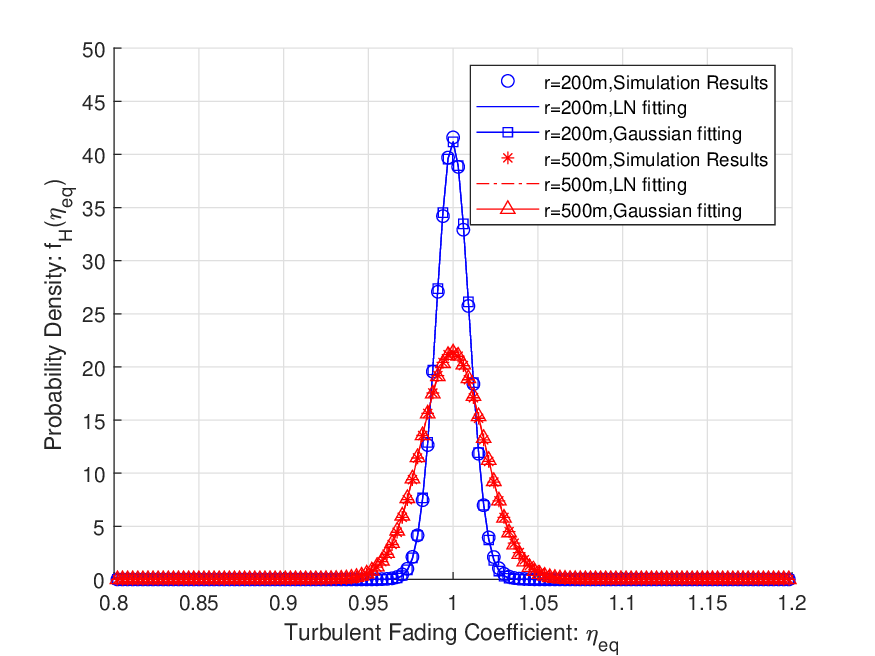}}
\subfigure[\label{PDF_moderate_tur}]
{\includegraphics[width=0.45\textwidth]{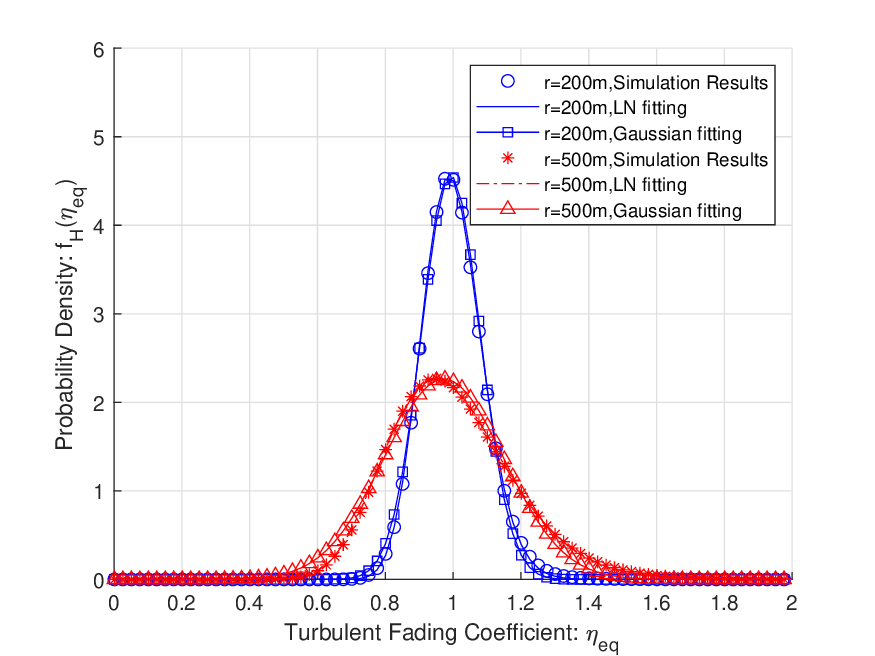}}
\subfigure[\label{PDF_strong_tur}]
{\includegraphics[width=0.45\textwidth]{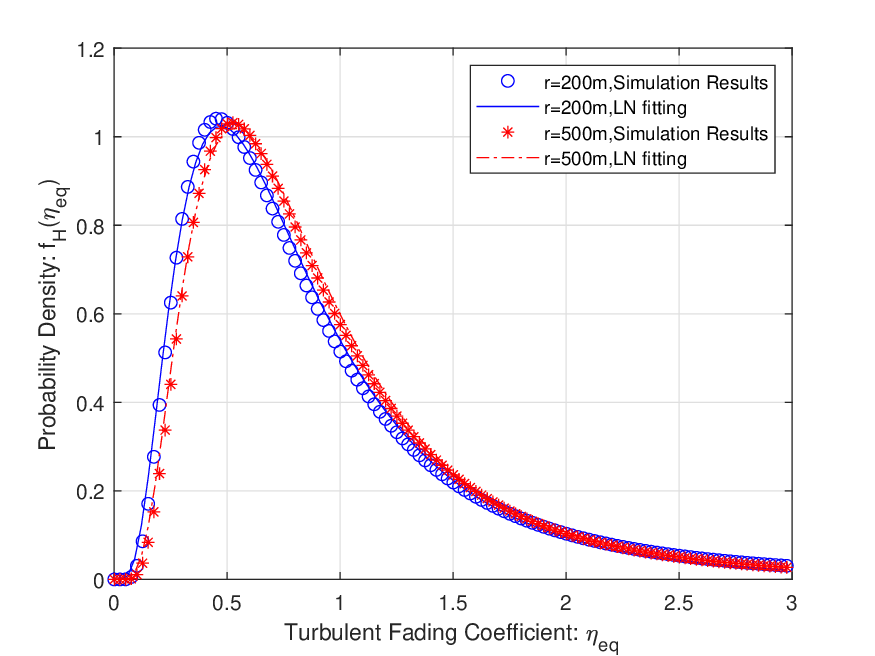}}
\subfigure[\label{PDF_compare}]
{\includegraphics[width=0.45\textwidth]{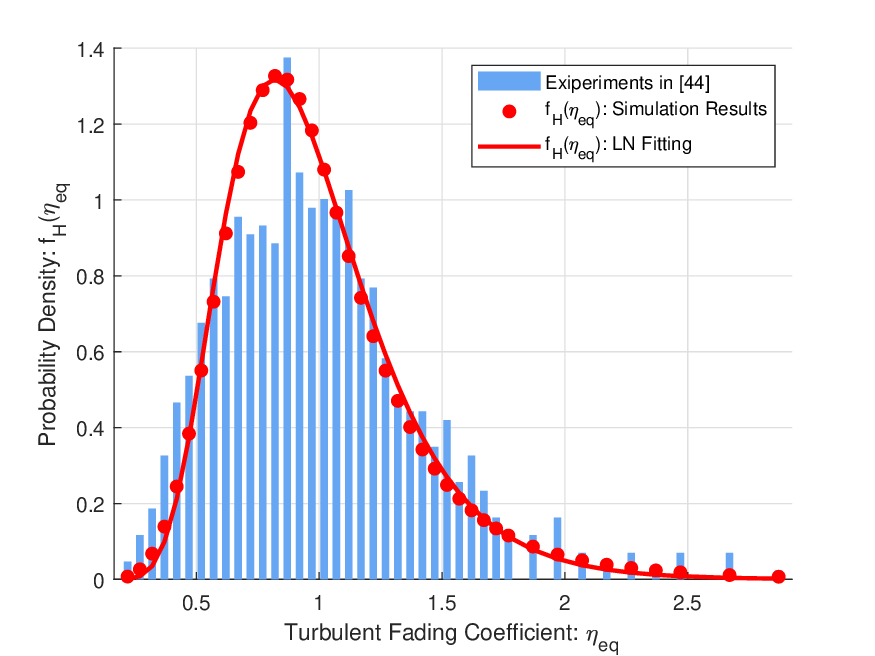}}
\caption{Distribution of turbulent fading coefficient under different turbulent conditions with $r=200$ m and $r=500$ m: (a) weak turbulence; (b) moderate turbulence; (c) strong turbulence; (d) PDF comparison with \cite{liao2014turbulence}}
\label{fig:PDF_comparison}
\end{figure*}

Then we present the estimated PDF for the turbulent fading coefficients under different communication distances $r=200$ m and $r=500$ m with different turbulent conditions in Fig. \ref{fig:PDF_comparison}. From Figs. \ref{PDF_weak_tur} to \ref{PDF_strong_tur}, we can observe that the simulation results of $f_{H}(\eta_{eq})$ can be well approximated by a LN distribution in all turbulent conditions. Specifically, when the turbulence is weak, a Gaussian distribution can also well fit the obtained PDF, which is reasonable because an LN distribution will approach a Gaussian distribution when the variance is small. Besides, we also compare our model with the turbulent distribution reported in an experimental work \cite{liao2014turbulence} in Fig. \ref{PDF_compare}. From Fig. \ref{PDF_compare} we can see that our model can well approximate the experimental results reported in  \cite{liao2014turbulence}.

\subsection{Experimental Results}

\begin{figure}
\centering
\subfigure[\label{Experimental_Platform}]
{\includegraphics[width=0.5\textwidth]{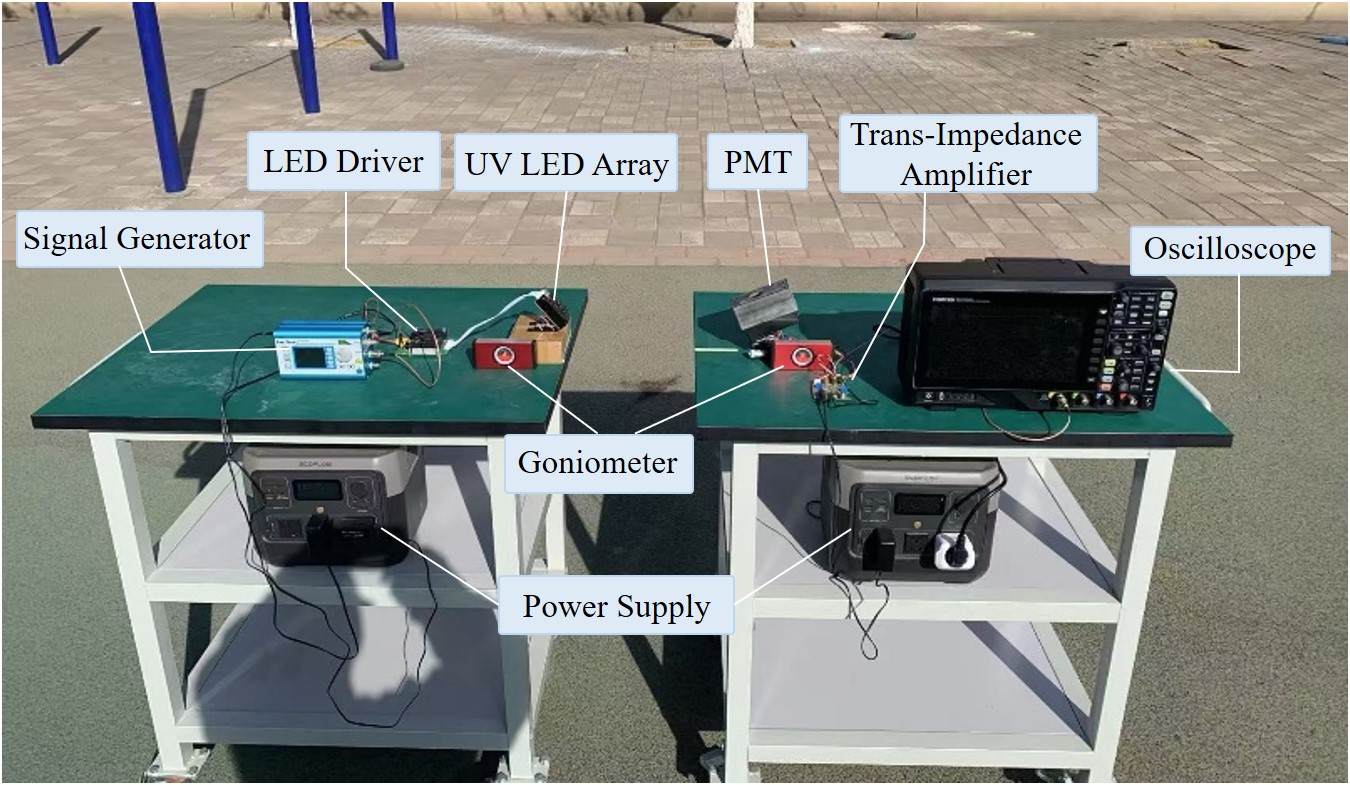}}
\subfigure[\label{50m_50deg_Voltage}]
{\includegraphics[width=0.24\textwidth]{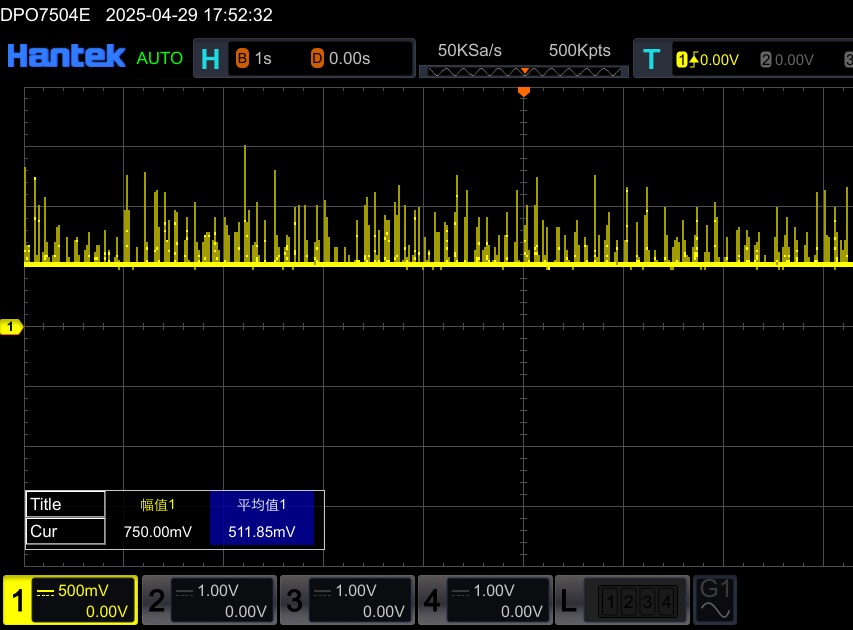}}
\subfigure[\label{50m_60deg_Voltage}]
{\includegraphics[width=0.24\textwidth]{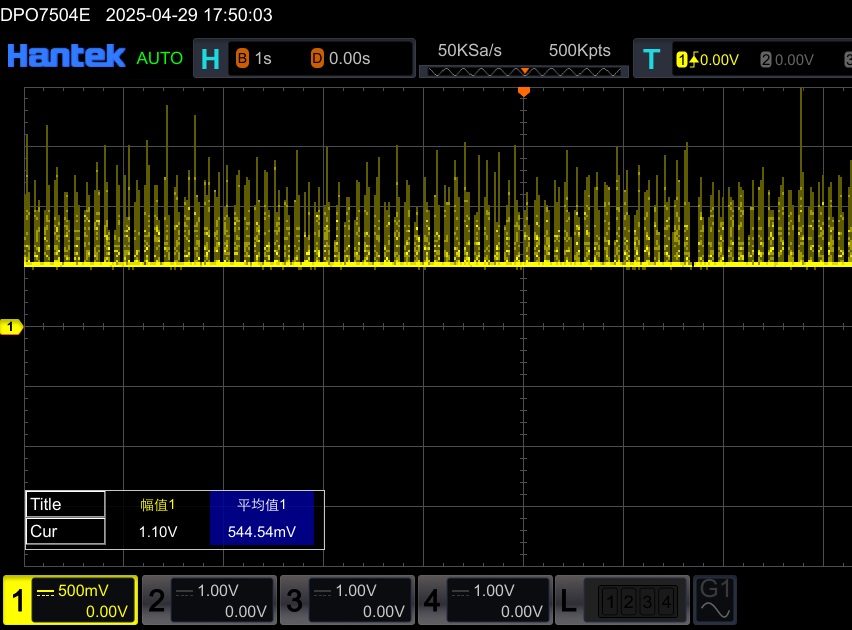}}
\caption{Experimental platform and typical receiving voltage under different zenith angles: (a) experimental platform; (b) $r=50$ m and $\theta_T=\theta_R=50^\circ$; (c) $r=50$ m and $\theta_T=\theta_R=60^\circ$}\label{fig: receiving_voltage}
\end{figure}

We established a experimental platform for measuring the turbulent fluctuation, shown in Fig. \ref{Experimental_Platform}. At the transmitter end, a signal generator is employed to generater on-off keying signals and the signal is directed to a LED driver. Then the LED driver drives the UV LED array and transmitting UV signals. The LED array consists of 8 LEDs with transmitting light power up to 30 mW and a divergence angle around $10^\circ$ (half-power divergence angle of Lambertian light source) for each LED. At the receiver end, a photomultiplier tube (PMT) combined with a UV filter is used to detect the scattered UV signal, then the detected signal is amplified by a trans-impedance amplifier and directed to an oscilloscope. The receiving FOV angle of the PMT is around $45^\circ$. The oscilloscope records the received signal and saves it for further processing. Two goniometers are employed to measure the transmitting and receiving zenith angles and two portable power suppliers are employed for power supply. The experiments were conducted on April 29, 2025 from 5:00 pm to 7:00 pm at Haidian campus of Beijing university of posts and telecommunications with ground wind speed around 4.5 $\text{m/s}$. The measurements were carried at two distances $r\in \{50 \text{m}, 80 \text{m}\}$ and six zenith angles with $\theta_T=\theta_R \in \{20^\circ, 30^\circ, 40^\circ, 50^\circ, 60^\circ, 70^\circ\}$. Two typical receiving signals in voltage under $(r=50 \text{m}, \theta_T=\theta_R=50^\circ)$ and $(r=50 \text{m}, \theta_T=\theta_R=60^\circ)$ are shown in Figs. \ref{50m_50deg_Voltage} and \ref{50m_60deg_Voltage}, respectively.

\begin{figure}
\centering
\subfigure[\label{Average_Voltage}]
{\includegraphics[width=0.45\textwidth]{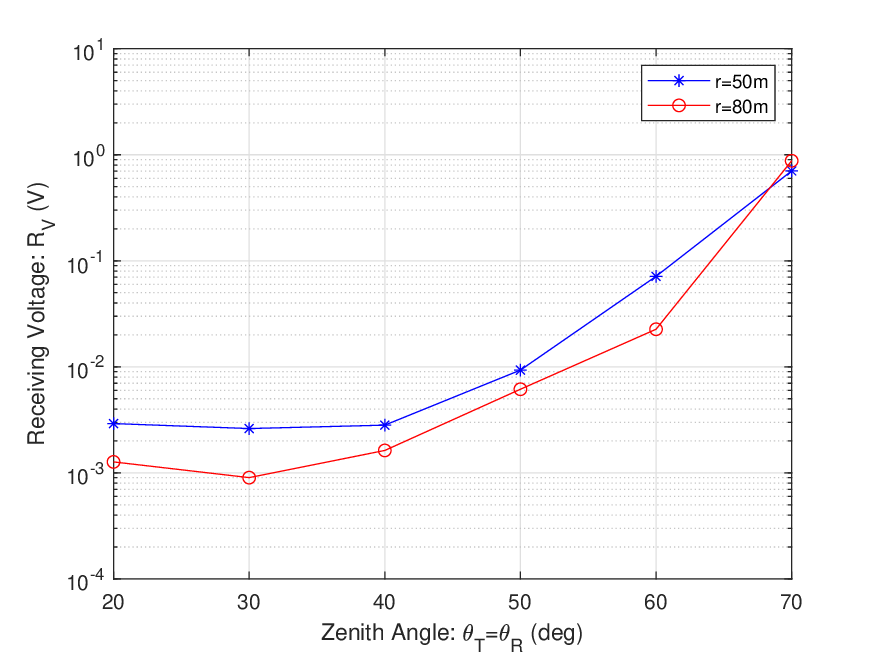}}
\subfigure[\label{Scintillation_Index}]
{\includegraphics[width=0.45\textwidth]{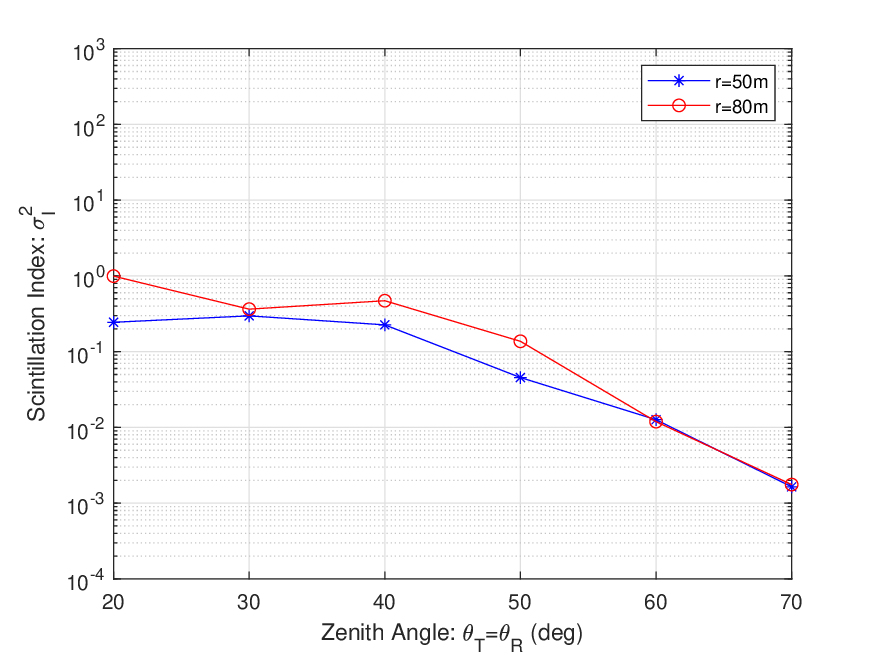}}
\caption{Experimental results of average receiving voltages and scintillation indices under different zenith angles: (a) Average receiving voltage; (b) Scintillation index}\label{fig: receiving_voltage_scintillation}
\end{figure}

Then we calculate the average receiving voltage and corresponding scintillation index under different transmitting distances and zenith angles. The scintillation index is calculated as $\sigma_I^2= \frac{\text{E}[R_V^2]}{\text{E}^2[R_V]}-1$, where $R_V$ is the receiving voltage. The obtained average voltage and scintillation index are shown in Figs. \ref{Average_Voltage} and \ref{Scintillation_Index}, respectively. From Fig. \ref{Average_Voltage} we can observe that a larger zenith angle corresponds to a larger average receiving voltage. Besides, a larger communication distance generally corresponds a smaller average receiving voltage. From Fig. \ref{Scintillation_Index}, we can see that the scintillation index decreases as the zenith angle increases. Besides, the scintillation index generally increases as the communication distance increases. These experimental results verified the simulation results in Fig. \ref{fig: var_tur_vs_geometry}. We also notice that the obtained data points demonstrated mess results when the zenith angle is too large. This is because when the zenith angle is too large, there exits small line-of-sight signals, which may affect the receiving signals.

\begin{figure}
\centering
\subfigure[\label{Gaussian_Fitting_50m}]
{\includegraphics[width=0.45\textwidth]{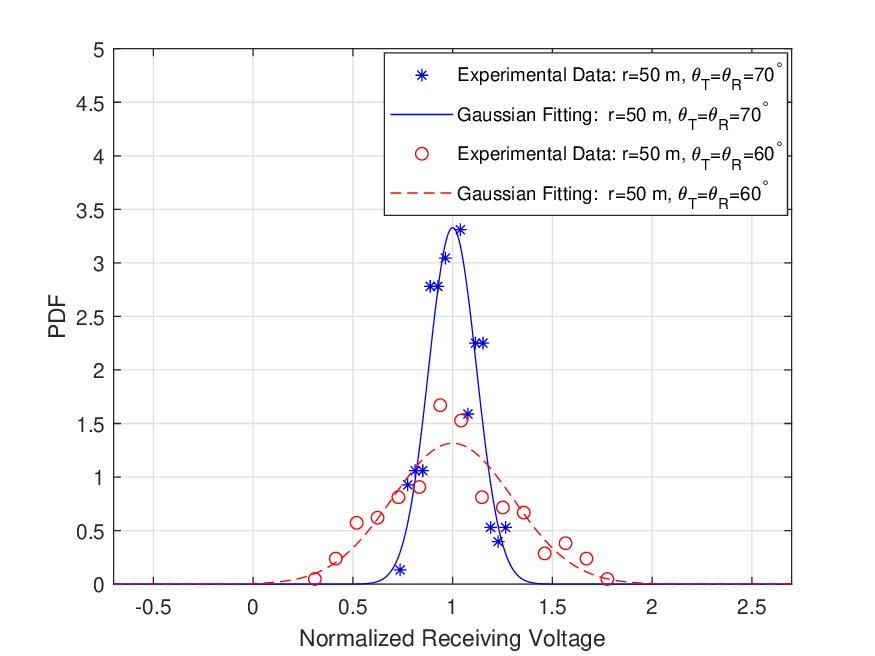}}
\subfigure[\label{Gaussian_Fitting_80m}]
{\includegraphics[width=0.45\textwidth]{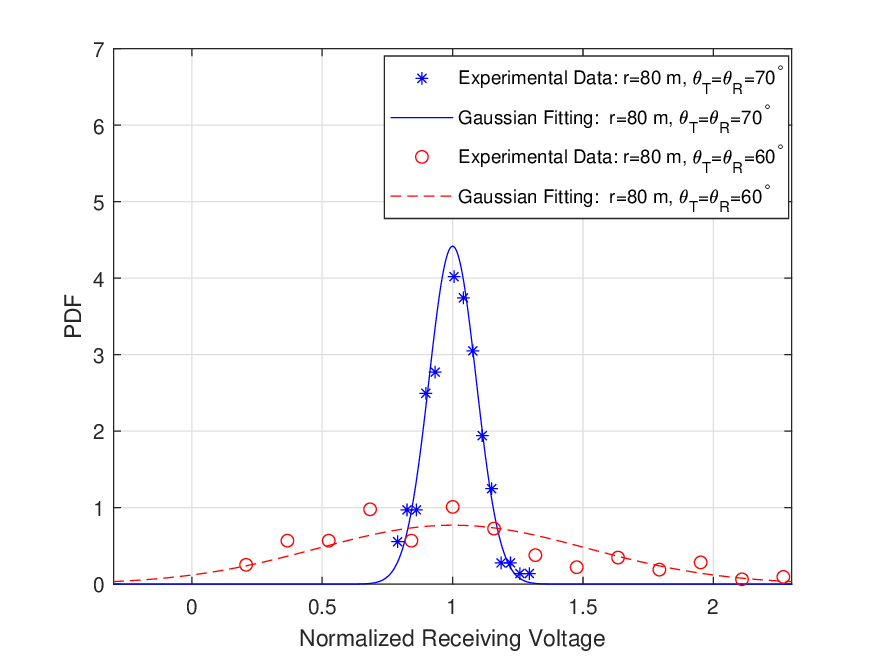}}
\caption{Experimental data and corresponding Gaussian fitting results under different distances: (a) $r=50$ m; (b) $r=80$ m}\label{fig: Gaussian_fitting}
\end{figure}

At last, we present the obtained distribution results from the normalized experimental data and the corresponding Gaussian fitting results under different communication distances and zenith angles in Fig. \ref{fig: Gaussian_fitting}. From Fig. \ref{fig: Gaussian_fitting} we can see that though the data points is not enough, we can also observe that a Gaussian fitting can fit the distribution of receiving signals under different communication distances and zenith angles. Because our experiments are conducted in relatively short range, a weak turbulence condition can be assumed, which verified our simulation results in Fig. \ref{PDF_weak_tur}.

\section{Conclusion}\label{Conclusion}

Existing works on the modeling of turbulent channels for NLOS UV communications either focused on the single-scattering cases or estimated the turbulent fluctuation effect in an unreliable way based on an MCS approach. In this paper, we established a turbulent multiple-scattering channel model by using an MCI approach, which is more efficient and interpretable compared with the MCS approach. Numerical results demonstrated that, for typical UV communication scenarios, the turbulence-induced scattering effect can always be ignored; and the turbulent fluctuation will increase as either the communication distance increases or the zenith angle decreases, which is compatible with existing experimental results and also with our experimental results. Besides, we demonstrated numerically that the distribution of the overall turbulent fading coefficient for the multiple-scattering process can be approximated as LN distributions under various turbulent conditions; and we also demonstrated both numerically and experimentally that the turbulent fading can be approximated as a Gaussian distribution under weak turbulence.

Our work can help researchers to accurately estimate communication performance, e.g., BER and outage probability, in turbulent channels, which is important for designing their UV communication system components, e.g., modulation schemes and link budget. For example, according to our findings, a more robust modulation scheme such as pulse-position modulation scheme rather than pulse-amplitude modulation scheme is preferred in large communication distances or elevation angles to combat the large turbulent fading. Besides, a higher risk of link interruption happens in a larger communication distance or elevation angle, and therefore requires a larger link budget in designing the communication system.

We have to point out that our experimental validation was conducted in an urban environment under relatively weak turbulence conditions. Turbulence characteristics can vary significantly across different scenarios due to variations in temperature gradients, humidity levels, and atmospheric composition. Though the effects of various weather conditions on UV NLOS communication was investigated in one of our previous work \cite{shan2023non}, the combination between the turbulent effects and the weather conditions has not been studied yet. In future work, we will further explore turbulence effects in diverse potential UV NLOS communication applications, such as maritime conditions with high humidity, desert regions with strong thermal turbulence, and high-altitude scenarios with reduced atmospheric density. Investigating how these environmental variations influence turbulence-induced fading could provide a more comprehensive understanding of UV NLOS channel characteristics across a broader range of channel conditions.

\bibliographystyle{IEEEtran}
\bibliography{ref}

@article{yuan2016review,
  title={Review of ultraviolet non-line-of-sight communication},
  author={Yuan, Renzhi and Ma, Jianshe},
  journal={China Commun.},
  volume={13},
  number={6},
  pages={63--75},
  year={2016},
  month={June},
  publisher={IEEE}
}

@article{yuan2016integral,
  title={An Integral Model of Two-Order and Three-Order Scattering for Non-Line-of-Sight Ultraviolet Communication in a Narrow Beam Case},
  author={Yuan, Renzhi and Ma, Jianshe and Su, Ping and He, Zehao},
  journal={IEEE Commun. Lett.},
  volume={20},
  number={12},
  pages={2366--2369},
  year={2016},
  month={Sep.},
  publisher={IEEE}
}

@inproceedings{yuan2019importance,
  title={An Importance Sampling Method for {Monte-Carlo} Integration Model for Ultraviolet Communication},
  author={Yuan, Renzhi and Ma, Jianshe and Su, Ping and Dong, Yuhan and Cheng, Julian},
  booktitle={2019 International Conference on Advanced Communication Technologies and Networking (CommNet)},
  pages={1--6},
  year={2019},
  month={Apr.},
  organization={IEEE}
}

@article{yuan2019monte,
  title={{Monte-Carlo} integration models for multiple scattering based optical wireless communication},
  author={Yuan, Renzhi and Ma, Jianshe and Su, Ping and Dong, Yuhan and Cheng, Julian},
  journal={IEEE Trans. Commun.},
  volume={68},
  number={1},
  pages={334--348},
  year={2019},
  month={Nov.},
  publisher={IEEE}
}

@article{wu2019single,
  title={Single-Scatter Model for Short-Range Ultraviolet Communication in a Narrow Beam Case},
  author={Wu, Tianfeng and Ma, Jianshe and Yuan, Renzhi and Su, Ping and Cheng, Julian},
  journal={IEEE Photonic. Tech. Lett.},
  volume={31},
  number={3},
  pages={265--268},
  year={2019},
  month={Feb.},
  publisher={IEEE}
}

@article{shen2021lmmse,
  title={{LMMSE-based SIMO} Receiver for Ultraviolet Scattering Communication With Nonlinear Conversion},
  author={Shen, Zanqiu and Ma, Jianshe and Su, Ping},
  journal={IEEE Wirel. Commun. Lett.},
  volume={10},
  number={10},
  pages={2140--2144},
  year={2021},
  month={Oct.},
  publisher={IEEE}
}

@article{yuan2023simo,
  title={Single-Input Multiple-Output Scattering Based Optical Communications Using Statical Combining in Turbulent Channels},
  author={Yuan, Renzhi and Peng, Mugen},
  journal={IEEE Trans. Wirel. Commun.},
  volume={23},
  number={4},
  pages={2560--2574},
  year={2024},
  month={Apr.},
  publisher={IEEE}
}

@article{wang2023nlos,
  title={Non-Line-of-Sight Full-Duplex Ultraviolet Communications Under Self-Interference},
  author={Wang, Zhifeng and Yuan, Renzhi and Peng, Mugen},
  journal={IEEE Trans. Wirel. Commun.},
  volume={22},
  number={11},
  pages={7775--7788},
  year={2023},
  month={Nov.},
  publisher={IEEE}
}

@article{wang2023mimo,
  title={MIMO Free-Space Optical Communications Using Photon-Counting Receivers Under Weak Links},
  author={Wang, Siming and Peng, Mugen and Yuan, Renzhi},
  journal={IEEE Commun. Lett.},
  volume={27},
  number={4},
  pages={1185--1189},
  year={2023},
  month={Feb.},
  publisher={IEEE}
}

@article{shan2020modeling,
  title={Modeling of ultraviolet omni-directional multiple scattering channel based on Monte Carlo method},
  author={Shan, Tao and Ma, Jianshe and Wu, Tianfeng and Shen, Zanqiu and Su, Ping},
  journal={Opt. Lett.},
  volume={45},
  number={20},
  pages={5724--5727},
  year={2020},
  month={Oct.},
  publisher={Optica Publishing Group}
}

@article{shan2020single,
  title={Single scattering turbulence model based on the division of effective scattering volume for ultraviolet communication},
  author={Shan, Tao and Ma, Jianshe and Wu, Tianfeng and Shen, Zanqiu and Su, Ping},
  journal={Chin. Opt. Lett.},
  volume={18},
  number={12},
  pages={120602},
  year={2020},
  month={Dec.},
  publisher={Chinese Optical Society}
}

@inproceedings{shen2020improved,
  title={Improved monte carlo integration models for ultraviolet communications},
  author={Shen, Zanqiu and Ma, Jianshe and Shan, Tao and Su, Ping},
  booktitle={2020 IEEE 20th International Conference on Communication Technology (ICCT)},
  pages={168--172},
  year={2020},
  organization={IEEE}
}

@article{shen2019modeling,
  title={Modeling of ultraviolet scattering propagation and its applicability analysis},
  author={Shen, Zanqiu and Ma, Jianshe and Shan, Tao and Wu, Tianfeng},
  journal={Opt. Lett.},
  volume={44},
  number={20},
  pages={4953--4956},
  year={2019},
  month={Oct.},
  publisher={Optica Publishing Group}
}

@article{chu2024turbulent,
  title={Turbulent single-scattering channel model for ultraviolet communications using equivalent scattering point approach},
  author={Chu, Xinyi and Yuan, Renzhi and Peng, Mugen},
  journal={IEEE Commun. Lett.},
  volume={44},
  number={20},
  pages={1579--1583},
  year={2024},
  month={July},
  publisher={IEEE}
}

@article{cao2023power,
  title={A Power-Domain MST Scheme With BPPM in NLOS Ultraviolet Communications},
  author={Cao, Tian and Wu, Tianfeng and Pan, Changyong and Song, Jian},
  journal={IEEE Photon. J.},
  volume={15},
  number={1},
  pages={1--10},
  year={2023},
  month={Feb.},
  publisher={IEEE}
}

@article{cao2023performance,
author={Cao, Tian and Wu, Tianfeng and Pan, Changyong and Song, Jian},
  journal={IEEE Commun. Lett.},
  title={Performance of Multipulse Pulse-Position Modulation in NLOS Ultraviolet Communications},
  year={2023},
  volume={27},
  number={3},
  pages={901-905},
  month={Mar.},
  publisher={IEEE}}

@article{cao2022single,
  title={Single-collision-induced path loss model of reflection-assisted non-line-of-sight ultraviolet communications},
  author={Cao, Tian and Wu, Tianfeng and Pan, Changyong and Song, Jian},
  journal={Opt. Express},
  volume={30},
  number={9},
  pages={15227--15237},
  year={2022},
  month={Apr.},
  publisher={Optica Publishing Group}
}

@article{cao2021single,
  title={Single-scatter path loss model of LED-based non-line-of-sight ultraviolet communications},
  author={Cao, Tian and Gao, Xinyu and Wu, Tianfeng and Pan, Changyong and Song, Jian},
  journal={Opt. Lett.},
  volume={46},
  number={16},
  pages={4013--4016},
  year={2021},
  month={Aug.},
  publisher={Optica Publishing Group}
}

@article{ding2009modeling,
  title={Modeling of non-line-of-sight ultraviolet scattering channels for communication},
  author={Ding, Haipeng and Chen, Gang and Majumdar, Arun K and Sadler, Brian M and Xu, Zhengyuan},
  journal={IEEE J. Sel. Areas Commun.},
  volume={27},
  number={9},
  pages={1535-1544},
  year={2009},
  month={Dec.},
  publisher={IEEE}
}

@article{drost2011uv,
  title={{UV} communications channel modeling incorporating multiple scattering interactions},
  author={Drost, Robert J and Moore, Terrence J and Sadler, Brian M},
  journal={J. Opt. Soc. Am. A},
  volume={28},
  number={4},
  pages={686--695},
  year={2011},
  month={Apr.},
  publisher={Optical Society of America}
}

@article{ding2010path,
  title={A path loss model for non-line-of-sight ultraviolet multiple scattering channels},
  author={Ding, Haipeng and Xu, Zhengyuan and Sadler, Brian M},
  journal={EURASIP J. Wirel. Commun. Netw.},
  volume={2010},
  pages = {63:1--63:11},
  year={2010},
  month={Apr.},
  publisher={Springer}
}

@article{gong2015channel,
  title={Channel estimation and signal detection for optical wireless scattering communication with inter-symbol interference},
  author={Gong, Chen and Xu, Zhengyuan},
  journal={IEEE Trans. Wirel. Commun.},
  volume={14},
  number={10},
  pages={5326--5337},
  year={2015},
  month={Oct.},
  publisher={IEEE}
}

@article{gong2016optical,
  title={Optical wireless scattering channel estimation for photon-counting and photomultiplier tube receivers},
  author={Gong, Chen and Zhang, Xiaoke and Xu, Zhengyuan and Hanzo, Lajos},
  journal={IEEE Trans. Commun.},
  volume={64},
  number={11},
  pages={4749--4763},
  year={2016},
  month={Nov.},
  publisher={IEEE}
}

@article{wei2018simultaneous,
  title={Simultaneous channel estimation and signal detection in wireless ultraviolet communications combating inter-symbol-interference},
  author={Wei, Zhuangkun and Hu, Wenxiu and Han, Dahai and Zhang, Min and Li, Bin and Zhao, Chenglin},
  journal={Opt. Express},
  volume={26},
  number={3},
  pages={3260--3270},
  year={2018},
  month={Feb.},
  publisher={Optical Society of America}
}

@article{hu2020non,
  title={Non-coherent detection for ultraviolet communications with inter-symbol interference},
  author={Hu, Wenxiu and Wei, Zhuangkun and Popov, Sergei and Leeson, Mark and Zhang, Min and Xu, Tianhua},
  journal={J. Light. Technol.},
  volume={38},
  number={17},
  pages={4699--4707},
  year={2020},
  month={Sept.},
  publisher={IEEE}
}

@article{qin2017noncoplanar,
  title={Noncoplanar geometry for mobile {NLOS MIMO} ultraviolet communication with linear complexity signal detection},
  author={Qin, Heng and Zuo, Yong and Li, Feiyu and Cong, Risheng and Meng, Lingchao and Wu, Jian},
  journal={IEEE Photon. J.},
  volume={9},
  number={5},
  pages={1--12},
  year={2017},
  month={Oct.},
  publisher={IEEE}
}

@inproceedings{he2024demonstration,
  title={Demonstration of a 900m Non-Line-of-Sight UV Communication System Using an LED Array},
  author={He, Nan and Shan, Tao and Cheng, Julian},
  booktitle={2024 IEEE/CIC International Conference on Communications in China (ICCC)},
  pages={132--137},
  year={2024},
  organization={IEEE}
}

@article{wang20181mbps,
  title={A {1Mbps} real-time {NLOS UV} scattering communication system with receiver diversity over 1km},
  author={Wang, Guanchu and Wang, Kun and Gong, Chen and Zou, Difan and Jiang, Zhimeng and Xu, Zhengyuan},
  journal={IEEE Photon. J.},
  volume={10},
  number={2},
  pages={1--13},
  year={2018},
  month={Apr.},
  publisher={IEEE}
}

@article{alkhazragi2020gbit,
  title={Gbit/s {ultraviolet-C} diffuse-line-of-sight communication based on probabilistically shaped DMT and diversity reception},
  author={Alkhazragi, Omar and Hu, Fangchen and Zou, Peng and Ha, Yinaer and Kang, Chun Hong and Mao, Yuan and Ng, Tien Khee and Chi, Nan and Ooi, Boon S},
  journal={Opt. Express},
  volume={28},
  number={7},
  pages={9111--9122},
  year={2020},
  month={Mar.},
  publisher={Optical Society of America}
}

@article{sun201771,
  title={{71-Mbit/s ultraviolet-B LED} communication link based on {8-QAM-OFDM} modulation},
  author={Sun, Xiaobin and Zhang, Zhenyu and Chaaban, Anas and Ng, Tien Khee and Shen, Chao and Chen, Rui and Yan, Jianchang and Sun, Haiding and Li, Xiaohang and Wang, Junxi and others},
  journal={Opt. Express},
  volume={25},
  number={19},
  pages={23267--23274},
  year={2017},
  month={Sept.},
  publisher={Optical Society of America}
}

@inproceedings{ding2011turbulence,
  title={Turbulence modeling for non-line-of-sight ultraviolet scattering channels},
  author={Ding, Haipeng and Chen, Gang and Majumdar, Arun K and Sadler, Brian M and Xu, Zhengyuan},
  booktitle={Atmospheric Propagation VIII},
  volume={8038},
  pages={195--202},
  year={2011},
  organization={SPIE}
}

@inproceedings{zuo2012effect,
  title={Effect of atmospheric turbulence on non-line-of-sight ultraviolet communications},
  author={Zuo, Yong and Xiao, Houfei and Wu, Jian and Hong, Xiaobin and Lin, Jintong},
  booktitle={2012 IEEE 23rd International Symposium on Personal, Indoor and Mobile Radio Communications-(PIMRC)},
  pages={1682--1686},
  year={2012},
  organization={IEEE}
}

@inproceedings{xiao2012non,
  title={Non-line-of-sight ultraviolet channel parameters estimation in turbulence atmosphere},
  author={Xiao, Houfei and Zuo, Yong and Fan, Cheng and Wu, Chaoye and Wu, Jian},
  booktitle={2012 Asia Communications and Photonics Conference (ACP)},
  pages={1--3},
  year={2012},
  organization={IEEE}
}

@article{xiao2013non,
  title={Non-line-of-sight ultraviolet single-scatter propagation model in random turbulent medium},
  author={Xiao, Houfei and Zuo, Yong and Wu, Jian and Li, Yan and Lin, Jintong},
  journal={Opt. Lett.},
  volume={38},
  number={17},
  pages={3366--3369},
  year={2013},
  month={Sept.},
  publisher={Optica Publishing Group}
}

@article{wang2013characteristics,
  title={Characteristics of ultraviolet scattering and turbulent channels},
  author={Wang, Peng and Xu, Zhengyuan},
  journal={Opt. Lett.},
  volume={38},
  number={15},
  pages={2773--2775},
  year={2013},
  month={Aug.},
  publisher={Optica Publishing Group}
}

@inproceedings{chen2014experimental,
  title={Experimental and simulated evaluation of long distance NLOS UV communication},
  author={Chen, Gang and Liao, Linchao and Li, Zening and Drost, Robert J and Sadler, Brian M},
  booktitle={2014 9th International Symposium on Communication Systems, Networks \& Digital Sign (CSNDSP)},
  pages={904--909},
  year={2014},
  organization={IEEE}
}

@inproceedings{liao2014turbulence,
  title={Turbulence channel test and analysis for NLOS UV communication},
  author={Liao, Linchao and Li, Zening and Lang, Tian and Sadler, Brian M and Chen, Gang},
  booktitle={Laser Communication and Propagation through the Atmosphere and Oceans III},
  volume={9224},
  pages={411--416},
  year={2014},
  organization={SPIE}
}

@article{liu2015performance,
  title={Performance analysis of non-line-of-sight ultraviolet communication through turbulence channel},
  author={Liu, Tao and Wang, Peng and Zhang, Hongming},
  journal={Chin. Opt. Lett.},
  volume={13},
  number={4},
  pages={040601--040601},
  year={2015},
  month={Apr.},
  publisher={Chinese Optical Society}
}

@article{xu2022improvement,
  title={Improvement of a Monte-Carlo-simulation-based turbulence-induced attenuation model for an underwater wireless optical communications channel},
  author={Xu, DongLing and Yue, Peng and Yi, Xiang and Liu, JingYi},
  journal={J. Opt. Soc. Am. A},
  volume={39},
  number={8},
  pages={1330--1342},
  year={2022},
  month={Aug.},
  publisher={Optica Publishing Group}
}

@article{shan2023non,
  title={Non-line-of-sight ultraviolet transmission coverage in non-precipitating, foggy, and rainy weather},
  author={Shan, Tao and Yuan, Renzhi and He, Nan and Cheng, Julian},
  journal={Opt. Express},
  volume={31},
  number={23},
  pages={37703--37721},
  year={2023},
  publisher={Optica Publishing Group}
}

@article{ardakani2017performance,
  title={Performance analysis of relay-assisted {NLOS} ultraviolet communications over turbulence channels},
  author={Ardakani, Maryam Haghighi and Heidarpour, Ali Reza and Uysal, Murat},
  journal={J. Opt. Commun. Netw.},
  volume={9},
  number={1},
  pages={109--118},
  year={2017},
  month={Sept.},
  publisher={Optica Publishing Group}
}

@article{gong2018full,
  title={On full-duplex relaying for optical wireless scattering communication with on-off keying modulation},
  author={Gong, Chen and Wang, Kun and Xu, Zhengyuan and Wang, Xiaodong},
  journal={IEEE Trans. Wirel. Commun.},
  volume={17},
  number={4},
  pages={2525--2538},
  year={2018},
  month={Apr.},
  publisher={IEEE}
}

@article{andrews2005laser,
  title={Laser beam propagation through random media},
  author={Andrews, Larry C and Phillips, Ronald L},
  journal={Laser Beam Propagation Through Random Media: Second Edition},
  year={2005}
}

@article{ishimaru2005theory,
  title={Theory and application of wave propagation and scattering in random media},
  author={Ishimaru, Akira},
  journal={Proceedings of the IEEE},
  volume={65},
  number={7},
  pages={1030--1061},
  year={2005},
  publisher={IEEE}
}

@book{kaushal2017free,
  title={Free space optical communication},
  author={Kaushal, Hemani and Jain, VK and Kar, Subrat},
  year={2017},
  publisher={Springer}
}

\end{document}